\documentclass[aps,prd,showpacs,superscriptaddress,groupedaddress,amsmath,nofootinbib,twocolumn]{revtex4}  
\usepackage{graphicx}  
\usepackage{dcolumn}   
\usepackage{bm}        
\usepackage{amssymb}   

\begin{document}
\renewcommand{\vec}[1]{\mathbf{#1}}
\newcommand{\df}[1]{\boldsymbol{#1}}

\widetext


\title{Classical Electrodynamics of Extended Bodies}
\affiliation{Colorado School of Mines, Golden, Colorado, USA}
\author{P.D.~Flammer} 
\email{pflammer@mines.edu}
\affiliation{Colorado School of Mines, Golden, Colorado, USA} 
\vskip 0.25cm

\date{\today}

\begin{abstract}
We study the classical electrodynamics of extended bodies. Currently, there is no self-consistent dynamical theory of such bodies in the literature. Electromagnetic energy-momentum is not conserved in the presence of charge and some addition is required. The only somewhat suitable addition found to date are point charges. These suffer from infinite self-energy, requiring some renormalization procedure, and perturbative methods to account for radiation. We review the history that has led to the understanding of these facts.

We then investigate possible self-consistent, non-point-charge, classical electrodynamic theories. We start with a Lagrangian consisting only of the Ricci scalar (gravity) and the standard electromagnetic field Lagrangian, and consider additions other than point charges and their associated interaction Lagrangian. Including quadratic terms in the Lagrangian involving first-order derivatives of the electromagnetic field tensor provides sufficient stress-energy terms to allow for conservation of energy-momentum. There are three such independent terms: a direct current-current interaction and two curvature-mediated (non-minimally coupled), short-range interactions, one of which changes sign under a parity transformation. These could be interpreted as non-electromagnetic, short-range forces.

For the simplest possible theory, with only the metric and the electromagnetic potential 1-form as independent fields, we find a single, stable, spherical (spin-0) solution, which due to an integrable singularity at the solution's center, has quantized mass and charge. Its charge is smaller than numeric error, and its mass is set by a new constant in the Lagrangian. It has a small, central core of charge surrounded by a wave of alternatingly charged, spherical shells, where the amplitude of the charge density wave is inversely proportional to the radial coordinate.


\end{abstract}

\maketitle

\tableofcontents

\section{\label{intro}Introduction}

There is currently no completely suitable theory describing the dynamics of charged bodies. Feynman, in his famous lectures, describes the situation as follows (see \cite{feynman1963} Vol.2 Ch. 28):
\begin{quote}
You can appreciate that there is a failure of all classical physics because of the quantum-mechanical effects. Classical mechanics is a mathematically consistent theory; it just doesn't agree with experience. It is interesting, though, that the classical theory of electromagnetism is an unsatisfactory theory all by itself. There are difficulties associated with the ideas of Maxwell's theory which are not solved by and not directly associated with quantum mechanics. You may say, ``Perhaps there's no use worrying about these difficulties. Since the quantum mechanics is going to change the laws of electrodynamics, we should wait to see what difficulties there are after the modification.'' However, when electromagnetism is joined to quantum mechanics, the difficulties remain. So it will not be a waste of our time now to look at what these difficulties are.
\end{quote}

The difficulties come in two flavors: (1) if a charged object is extended, i.e. it has some non-zero size, it has not been possible to develop self-consistent equations of motion for that object; and (2) if we take the point-charge limit (to resolve difficulty 1, or because we think physical charges are points), the energy of the charge becomes infinite. Feynman continues by discussing examples in the literature of researchers trying to resolve the infinite energy of point charges, finally concluding that all attempts have failed\cite{feynman1963}:
\begin{quote}
We do not know how to make a consistent theory-including the quantum mechanics-which does not produce an infinity for the self-energy of an electron, or any point charge. And at the same time, there is no satisfactory theory that describes a non-point charge. It's an unsolved problem.
\end{quote}

The infinite ``self-energy'' of a point charge is due to self-interaction: its constituent parts repel against the other parts; as one tries to pack charge into a ball, the smaller the ball, the harder this is, and more work must be done to make it more compact. The result is the energy required to form such a ball is inversely proportional to the size of the ball; and point charges have infinite self-energy. This is in apparent contradiction to the physical fact that a very compact electron exists.

Another consequence of self-interaction is radiation: as a charge is accelerated, its interaction with its own field (in addition to adding to its rest energy) can cause the charge to recoil, as momentum is carried away by the fields in the form of radiation. Again, the classical reaction of a charge to radiation is intractable. J.D.~Jackson summarizes this difficulty in his text as (See Sec.~16.1 of \cite{jackson}):
\begin{quote}
...a completely satisfactory classical treatment of the reactive effects of radiation does not exist. The difficulties presented by this problem touch one of the most fundamental aspects of physics, the nature of an elementary particle. Although partial solutions, workable within limited areas, can be given, the basic problem remains unsolved.
\end{quote}

As we will see in the next section, the root cause of these problems is our ignorance of what keeps an electron (or other compact charge) compact. As it turns out, it has been very difficult to even postulate a self-consistent theory of what could bind an electron (or other compact charge) together; no such theory exists to date.

In this paper, we attempt to address this fundamental question. This cannot be addressed by treating point charges, as they are compact by construction. Also, this cannot be done in a quantum mechanical framework, because quantum mechanics presupposes point charges (the idea of an extended object is foreign to the theory). Therefore, we investigate the classical electrodynamics of extended bodies.

\section{\label{history}Historical Review}

We start by reviewing some of the history of the development of the theory of electrodynamics. The main purpose of this section is to explore the evolution of thought, evolving from Coulomb's law to the theory of quantum electrodynamics, focusing on the issue of self-interaction, the resulting self-inconsistency of electrodynamic theory, and the necessary introduction of point charges along with their pathologies. 


\subsection{\label{maxwell}Development of Maxwell's Equations}

With the invention of the Leyden Jar (a rudimentary capacitor) in the middle of the 18$^{\rm th}$ century, experimentalists were able to repeatably apply charge to various objects, and determine how charged objects affect each other. By 1785, Coulomb had established the mathematical form of this electrostatic force\cite{coulomb1785}, the law being very similar to Newton's law of gravitation.

Near the turn of the 19$^{\rm th}$ century, Alessandro Volta invented the voltaic pile (battery). This enabled experimentalists to more reliably study electrical flow in circuits. In the summer of 1820, Oersted discovered the amazing fact that magnetic needles were affected by electric currents, linking what were before thought to be separate phenomena, electricity and magnetism\cite{jackson2001}. Within a few months, Biot and Savart successfully determined the mathematical behavior of the force between a current carrying wire and a magnetic pole\cite{biotsavart1820, biotsavart1820b, biot1824}. By 1827, Ampere had also shown that solenoids of current carrying wire behaved similarly to bar magnets, and studied the magnetic force between two circuits\cite{ampere1827}. 

Also in the 1820s, Ohm successfully described that the current in a conductor was proportional to the electromotive force and the conductance of the material\cite{ohm1827}. This was the primary ``force law'' (now called Ohm's law) used by physicists for electrodynamics until near the turn of the 20$^{\rm th}$ century. In 1831, Faraday discovered that moving a magnet near a wire circuit induced a current in the circuit, discovering electromagnetic induction\cite{faraday1939}.  

Various physicists worked to understand the interaction between magnets and currents for the next few decades\cite{neumann1847, grassmann1845}. One theoretical achievement, which was important to the development of electromagnetic theory, was the use of ``potentials''.  In 1857, Kirchhoff first wrote the electric force as a combination of the gradient of a scalar potential (which had already been used for some time in electrostatic problems) and the time derivative of a newly introduced vector potential\cite{kirchhoff1857}. Kirchhoff also showed, in that particular formulation, that the vector and scalar potential were related to one another (in modern terminology, describing the particular gauge, which he was using).

All of this work found some closure in the 1860s. In 1861 and 1862, Maxwell published ``On physical lines of force''\cite{maxwell1861} (where he added the necessary displacement current\footnote{The displacement current is mathematically necessary to conserve charge.}), and in 1865, he presented a complete framework of electromagnetism in ``A Dynamical Theory of the Electromagnetic Field''\cite{maxwell1865}. This theory was extremely successful at describing all of the electrical phenomena known at the time; he also calculated that electromagnetic waves propagate at a speed close to the speed of light (which had recently been measured), thus identifying light as an electromagnetic wave.

A key piece of this new theory was that important dynamics took place in the space between electrified objects. This was a major shift in thought: up to that time, interactions were typically thought of as ``actions at a distance''. In Maxwell's words:
\begin{quote}
These [old] theories assume, more or less explicitly, the existence of substances the particles of which have the property of acting on one another at a distance by attraction or repulsion.\cite{maxwell1865}
\end{quote}
Maxwell differentiated his new theory in this way:
\begin{quote}
The theory I propose may therefore be called a theory of the {\em Electromagnetic Field}, because it has to do with the space in the neighborhood of the electric or magnetic bodies.\cite{maxwell1865}
\end{quote}
This was the birth of physical field theories, where the original concept of a ``field'' was that important dynamics occur (and propagate) throughout the space (or field) between interacting bodies.

To motivate the fact that electromagnetic interactions could propagate through ``so-called vacuum'', Maxwell used the idea of disturbances propagating through an elastic medium, called the  ``luminiferous aether''. However, although Maxwell used this idea of an underlying elastic medium to develop the theory, he gave up on hypothesizing its exact character or role:
\begin{quote}
I have on a former occasion, attempted to describe a particular kind of motion and a particular kind of strain, so arranged as to account for the phenomena. In the present paper, I avoid any hypothesis of this kind; and in using words such as electric momentum and electric elasticity in reference to the known phenomena of the induction of currents and the polarization of dielectrics, I wish merely to direct the mind of the reader to the mechanical phenomena which will assist him in understanding the electrical ones. All such phrases in the present paper are to be considered as illustrative, not as explanatory.\cite{maxwell1865}
\end{quote}
Immediately after the previous statement, however, he stresses the importance of the field:
\begin{quote}
In speaking of the Energy of the field, however, I wish to be understood literally... On the old theories it resides in the electrified bodies, conducting circuits, and magnets... On our theory it resides in the electromagnetic field, in the space surrounding the electrified and magnetic bodies, as well as in those bodies...\cite{maxwell1865}
\end{quote}
The fact that the field could contain energy in its own right allowed him to effectively describe how fields transport energy via radiation through a vacuum, and equate light and heat with electromagnetic waves\footnote{Later in the 1860s, Lorenz and Riemann alternatively described the interactions between currents and charged objects as retarded integrals of the charge and current rather than focusing on the dynamics of the field\cite{lorenz1867,riemann1867}. This point of view had some advantages; in particular, it didn't motivate the existence of the aether. However, in the late 19$^{\rm th}$ and early 20$^{\rm th}$ centuries, electromagnetic theory predominantly grew out of Maxwell's theory, and the contributions of Lorenz and Riemann were somewhat forgotten until later\cite{jackson2001}}.

\subsection{\label{discreteRadiation}Radiation, Self-Interaction, and Special Relativity}

With the connection of electromagnetism and light, it became clear that currents, which change in time, generate electromagnetic waves, i.e. radiation. The radiated energy due to a varying electrical current was calculated by Fitzgerald in 1883\cite{fitzgerald1883}, and a general vectorial law for the flow of electromagnetic energy and its conservation was derived in 1884 by Poynting\cite{poynting1884}. Experimental generation and measurement of electromagnetic radiation at lower-than-optical frequencies was achieved by Hertz in 1887 using oscillating electrical circuits\cite{hertz1892,jackson2001}. Poincar\'{e} immediately noted that such radiation must cause damping within the oscillator due to the energy it carries away\cite{poincare1891}.


Also in 1887, the concept of the aether was discounted by the experiment of Michelson and Morley\footnote{This showed that the speed of light was independent of direction; very unlikely if it is a disturbance of an underlying medium that the earth was likely moving through.}. With the aether gone (or at least going), the ``field'' was no longer a description of ``space in the neighborhood''. The electromagnetic field necessarily took on a life of its own; the field (or radiation, or the energy-momentum it carries, etc.) was its own substance.

Up until the late 19$^{\rm th}$ century, experimentally and theoretically, continuous charge and current densities (in circuits) were the primary focus of study: Maxwell's equations were used to calculate the field, and Ohm's law was used to calculate how fields caused the current to evolve. However, in the 1880s, many researchers turned their attention to calculating the fields of discrete, compact charges, including Heaviside\cite{heaviside1889}, who is often credited with writing Maxwell's equations in their more modern form.

In 1892, Lorentz published ``Maxwell's electromagnetic theory and its application to moving bodies''; in this paper, he wrote down the force from an external electromagnetic field on a charged particle (point charge), now called the Lorentz force; he also formalized the gauge invariance of electromagnetism\cite{lorentz1892}. Lorentz also noted that, in general, one must account for the electromagnetic force of a discrete charge on itself. In his 1892 paper, Lorentz evaluated this self-force and calculated the equations of motion for a ``relativistically rigid'' spherical shell of charge (where the sphere maintains its shape in its proper frame)\footnote{Lorentz called this model a deformable sphere, because he noticed (before Einstein's theory of relativity) in a moving frame, the electron would contract, but this model is now called relativistically rigid\cite{yaghjian}.}. This was done in the limit of the sphere being small, so higher order terms in the size of the sphere could be ignored. It was found that the self-force contains a term, with magnitude inversely proportional to the size of the sphere, which can effectively be added to its inertial mass.


Additionally, a term appeared in the force equation, which is independent of the size. This force came to be known as the ``radiation reaction'' or ``field reaction'', although it seems Lorentz initially did not connect this reaction to radiation; Planck appears to be the first to do so in 1897\cite{planck1897}. Also in 1897, J.J. Thomson discovered the existence of the electron, which fueled further study of small, discrete charges.

Lorentz initially calculated this self-field reaction in the low-velocity limit (or, if you like, in the proper frame of the charged body), but by the early 1900s, Abraham (and then Lorentz) had extended this theory to arbitrary velocities\cite{abraham1903,lorentz1904}. Also, during this time, the hypothesis that the electron mass was due entirely to the electromagnetic self-interaction gained some favor (Abraham explicitly assumed it was the only contributor to the electron mass\footnote{It appears this was done, at least in part, because at the time, it was thought (before Einstein's Special Relativity) that any other mass would not transform between reference frames in the same way as electromagnetic mass; see \cite{yaghjian} for a discussion of this history.}).

In 1905, Einstein published ``On the Electrodynamics of moving bodies'', where he introduced his concept of special relativity\cite{einstein1905a}. In a paper later that year, he proposed that the inertial mass of a body was directly proportional to its energy content\cite{einstein1905b}. With this, one could calculate the mass due to the energy stored in the electromagnetic field for a charged object. It's interesting to note that, although they preceded special relativity, the equations of Lorentz and Abraham exhibited many special-relativistic effects (e.g. the fact that the speed of the charge can only asymptotically approach the speed of light).

\subsection{\label{sec:issues}Issues with Self-Interaction}

All of these developments gave some hope that a fully successful model of the electron was within reach. However, there were serious issues with the model. In 1904, Abraham derived a power equation of motion for the rigid model of an electron. Unfortunately, the power equation was not consistent with the force equation derived earlier: the scalar product of the velocity and the force does not equal the power. Also, in the context of relativity, the power and the force do not form a 4-vector.

There is an issue with the inertial mass as well, which one calculates from the Lorentz-Abraham equations: for a spherical shell, it is 4/3 times the mass that one obtains from the energy stored in the electrostatic fields (the self-energy). This was not noticed originally by Lorentz or Abraham as their theory preceded special relativity, but in the second edition of Abraham's book, Abraham mentions this discrepancy\cite{abraham1905, yaghjian}.

The equations of motion also violate causality. If a force is instantaneously ``turned on'' and one excludes runaway solutions, pre-acceleration solutions exist (the charge accelerates before the force is turned on)\cite{yaghjian}. 

In 1906, Poincar\'{e} pointed out the source of most of these problems: in order for a stable charged object to exist, there must be non-electromagnetic forces, which bind the electron together (keeping it from exploding due to its self-electric field): ``Therefore it is indeed necessary to assume that in addition to electromagnetic forces, there are other forces or bonds''\cite{poincare1906,yaghjian}. He came to the conclusion that while this other binding force integrated to zero over the object, the integrated power from the binding force was not zero, and exactly canceled the discrepancy between the force and power equations. However, this did not resolve the ``4/3 problem''. In order to correct that, one must include some ``bare mass'' of the charge\footnote{The bare mass of a charge is what its mass would be if it had no electromagnetic field. One cannot set this to zero for arbitrary geometries of charge. If one sets the geometry, the bare mass must take a specific value in order to be consistent with the self-energy\cite{yaghjian, flammer2016}}, which was set to zero by early authors.

The problems associated with the radiation reaction, the 4/3 problem, pre-acceleration, etc., continue to receive some attention in the literature. See the following references for examples from the 21$^{\rm st}$ century\cite{medina2006,rohrlich2008,aguirregabiria2006,essen2015,ferris2011,heras2003,rohrlich2000,rohrlich2001,steane2015,steane2015b,steane2015c,villarroel2002}. Ref.~\cite{rohrlich1997} has a concise historical overview of the problem.

A full history and detailed treatment of the spherical shell, with a description of the cause of these paradoxes may be found in a comprehensive monograph by Arthur Yaghjian\cite{yaghjian}. In particular it's worth noting, the pre-acceleration issue can be traced to the fact that ``turning on'' a force creates a non-analytic point in the force as a function of time, which invalidates the derivation of the equations of motion. If the force is analytic as a function of time, no pre-acceleration appears in the point-charge limit\cite{yaghjian}. The remainder of the problems are due to two missing items in electromagnetic theory: the bare mass of electromagnetic charge and a suitable binding force to bind a compact charge together.

It is an interesting fact of relativity and electrodynamics, that one has less freedom in choosing the problems one can treat than in general classical mechanics, such as instantaneously ``turning on'' a force (although conceptually possible, it's difficult to think a real relativistically valid force can turn on instantly). Most strikingly, one cannot consistently consider the dynamics of a charged object without appropriately balancing the forces on its constituent parts and knowing its internal mass density. The idea of considering a blob of charge, without considering what it is made of and what binds it together, yields inconsistent equations of motion.


One may attempt to model a certain structure, like the rigid sphere, and after the fact, add in what the binding force must have been, and state what it's bare mass must have been. However, this inherently violates causality, since the binding force is required to react across the entire object instantaneously. 

Thus, in order to really create a self-consistent dynamical theory for extended charged bodies, one must know two things {\em a priori}: the local binding force density that creates stability, and the bare mass density of the charge. There was some effort in the early 20$^{\rm th}$ century to this end. In the 1910s, an idea originated by Mie generated some hope, albeit short-lived\cite{mie1912,weyl1918}. Einstein commented on these developments in 1919:
\begin{quote}
Great pains have been taken to elaborate a theory which will account for the equilibrium of the electricity constituting the electron. G. Mie, in particular, has devoted deep researches to this question. His theory, which has found considerable support among theoretical physicists, is based mainly on the introduction into the energy-tensor of supplementary terms depending on the components of the electro-dynamic potential, in addition to the energy terms of the Maxwell-Lorentz theory. These new terms, which in outside space are unimportant, are nevertheless effective in the interior of the electrons in maintaining equilibrium against the electric forces of repulsion. In spite of the beauty of the formal structure of this theory, as erected by Mie, Hilbert, and Weyl, its physical results have hitherto been unsatisfactory. On the one hand the multiplicity of possibilities is discouraging, and on the other hand those additional terms have not as yet allowed themselves to be framed in such a simple form that the solution could be satisfactory\cite{einstein1919}.
\end{quote}
In the same paper, Einstein proposed gravity as a possible binding force for the electron by modifying his field equations; this admitted stable solutions, but could not explain charge quantization, causing him to abandon that line of thought\cite{einstein1919}.

None of these studies came to result in any suitable theory, and eventually support for this direction waned. To quote Weyl from the early 1920s,
\begin{quote}
Meanwhile I have quite abandoned these hopes, raised by Mie's theory; I do not believe that the problem of matter is to be solved by a mere field theory\cite{weyl1918}.
\end{quote}

\subsection{\label{pointChargeQM}Point Charges and Quantum Mechanics}

Without knowing the bare mass density and binding force for a charged object, one cannot solve for what stable objects should exist. But as mentioned in the last section, one can assume a structure, and add in what the binding force and bare mass should have been after the fact, unfortunately violating causality since the force is required to react instantaneously across the object (to maintain the pre-ordained shape). But if the object is a point, no time is required for signals to cross the object and causality is restored\footnote{The external force still must be analytic as a function of time, or you will still have the pre-acceleration issues discussed earlier\cite{yaghjian}.}. Also, this removes all internal degrees of freedom, so conservation of total momentum determines the motion completely.

One cannot directly calculate the energy or mass of an object, because we do not know the bare mass. What we can calculate, from the electromagnetic energy of the charge, diverges in the point-charge limit\cite{heaviside1889}. But the mass of the electron is a measurable quantity; rather than calculate it, one may simply use its measured value. This process of replacing an infinite calculated value with a measured value is often called ``renormalization''. In the context of classical dynamics of an electron, Dirac is credited with writing down the ``renormalized'' classical equations of motion of a charged particle in 1938\cite{dirac1938}, where he developed these equations in a manifestly covariant method\footnote{Note Von Laue had already written down the covariant radiation reaction much earlier\cite{vonlaue1909}.}. These renormalized equations of motion are often called the Lorentz-Abraham-Dirac equations of motion.

In any case, in the early 1900s, atomic structure was forcing physicists to rethink their perception of reality. With the discovery of the atomic nucleus around 1910, the idea that electrons orbit the nucleus (in the same way as planets orbit the sun) took root. However, any simple classical model of the electron (such as Lorentz's sphere of charge) cannot produce stable orbits around an atomic nucleus, precisely due to the damping effect of the radiation reaction: an electron in orbital motion will radiate energy away and its orbit will decay. If one ignores the radiation reaction, then classically one finds a continuum of possible orbits, which is also not what is measured: discrete, stable energy levels are observed for atomic orbits.

Due to these difficulties, in the 1910s and 1920s, a new way of thinking about these physical systems emerged, which were more successful at describing atomic phenomena: quantum mechanics. In the ``old quantum mechanics'' (sometimes called the Bohr model, or Bohr-Sommerfeld model), there was not much departure from classical thought. The electron was assumed to exist as a point (or at least very small) charge; classical orbits were then solved for the electron, and integrals of generalized momenta along the orbits were required to be integer multiples of the Planck constant, which yielded the correct energy levels (see Sommerfeld's 1921 book on the subject\cite{sommerfeld1921}). Note however, the classical equations of motion, which were used, ignored the radiation reaction, and the topic of the stability or self-interaction of particles was avoided altogether. This was very successful at predicting energy levels for simple systems, such as Hydrogen.

In the last half of the 1920s, the more modern quantum mechanics took shape. A new ``wave mechanics'' approach was developed where classical equations of motion, such as the classical Hamiltonian (again without any self-interaction/radiation reaction), are taken, and ``quantized'' (dynamic variables become operators on a wave function, which describes the state of the system) to develop equations such as the Schrodinger equation, which was published in 1926\cite{schrodinger1926}. At the same time, a different formulation, ``matrix mechanics'' was developed by Heisenberg, Born, and Jordan\cite{born1925}, which was shown to be equivalent to the wave mechanics approach.

Initially, all of this was done for low, non-relativistic velocities. However, in 1930, Dirac developed the relativistic generalization of Schrodinger's equation for electrons\cite{dirac1930}. With the success of the Dirac equation in predicting energy levels in simple atoms (including the interaction with the electron spin), attention turned to describing self-interaction/radiative corrections in the framework of quantum mechanics. This was done by starting with a non-interacting solution, and ``perturbing'' it by adding in successive interaction terms (these days, ``Feynman diagrams'' are used to do bookkeeping on what terms are needed)\cite{feynman1949}. However, any attempt to add in certain self-interaction terms resulted in infinity (similar to the classical case). In 1945, Feynman and Wheeler published ``Interaction with the absorber as the mechanism of radiation'', where they proposed that electrons do not interact with themselves at all\cite{wheeler1945}, illustrating some of the frustration of the time. (see \cite{feynman1963} Vol. 2, Ch. 28 for more discussion and other examples of efforts to remove this infinity). But by 1949, Schwinger and Tomonaga developed methods that circumvent the issue of infinite self-interaction, while accurately predicting the Lamb shift and the anomalous magnetic moment of the electron. Infinite self-interaction terms are absorbed into quantities, such as mass and charge, and the experimentally measured values are used in place of the infinite calculated ones\cite{schwinger1948,feynman1949}. As mentioned above, this process of dealing with infinite calculated values is called renormalization\footnote{Often in practice, one assumes a bare mass or bare charge that are also infinity in just the way needed to cancel the infinite self-interaction and result in the measured value.}. The perturbative process of adding in appropriate interaction terms, in conjunction with renormalization is what is now called quantum electrodynamics.

The ``standard model'', built on these principles, is extremely successful at predicting quantities outside of those which require renormalization. While the renormalization program allows physicists to do useful calculations, the lack of ability to calculate the masses of particles is less than ideal. In 1979, Dirac, speaking of renormalization, said
\begin{quote}
It's just a stop-gap procedure. There must be some fundamental change in our ideas, probably a change just as fundamental as the passage from Bohr's orbit theory to quantum mechanics. When you get a number turning out to be infinite which ought to be finite, you should admit that there is something wrong with your equations, and not hope that you can get a good theory just by doctoring up that number.\cite{buckley1979} 
\end{quote}
Feynman, who shared a Nobel prize, in part for developing the renormalization program, also was skeptical in his later years. In his 1986 book, he wrote
\begin{quote}
The shell game that we play to find $n$ [bare mass] and $j$ [bare charge] is technically called ``renormalization.'' But no matter how clever the word, it is what I would call a dippy process! Having to resort to such hocus-pocus has prevented us from proving that the theory of quantum electrodynamics is mathematically self-consistent. It's surprising that the theory still hasn't been proved self-consistent one way or the other by now; I suspect that renormalization is not mathematically legitimate.\cite{feynman1986}\footnote{It is interesting that some, who were so integral to developing quantum electrodynamics into its current state, had such opinions; it may be the only time someone has described a good portion of their Nobel Prize winning topic as ``dippy''.}
\end{quote}


In addition to failing to predict quantities such as mass and charge of particles, renormalization is somewhat at odds with general relativity: general relativity is non-renormalizable (one cannot play the same game and get any useful predictions from calculations). This troubling fact is a motivator in the study of string theory, where particles are stretched into strings: different excitations of strings are the different particles of nature, with finite self-energies. Unfortunately, to this date, despite significant effort, string theory has yet to demonstrate itself as a suitable theory that can predict experimental results. 

\subsection{Gravity and Electromagnetism}

Speaking of general relativity, we skipped over some details on historical attempts to integrate the theory of gravitation and electromagnetism. Einstein introduced his theory of gravitation, general relativity, in 1915\cite{einstein1915}.  However, since gravity is most important at astronomical scales, for planets, stars, black holes, etc., which are not likely to carry significant excess charge, the vast majority of theoretical and computational studies in  general relativity consider uncharged situations\cite{ivanov2002}.

There have been various efforts throughout the 20$^{\rm th}$ century to ``unify'' gravity and electromagnetism, where researchers have attempted to describe electromagnetism in the context of a generalized theory of the geometry of space-time (see \cite{goenner2004} for a review). However, more discussion of this type of unification of gravity with electromagnetism does not contribute to our purpose here. For the entirety of this paper we take a conventional ``dualistic view'', where matter is treated separate from geometry; it is the source of geometry's curvature.

Concerning the presence of electromagnetic charge in conventional general relativity: the metric for the space outside of a charged spherical object was published in 1916 and 1918 by Reissner\cite{reissner1916} and Nordstrom\cite{nordstrom1918}. Study of the interior of charged objects was not attempted until more recently than other history outlined here, starting mainly in the latter half of the 20$^{\rm th}$ century. For example, charged polytropic stars have been studied\cite{ray2004}, as well as charged situations with various other equations of state and space-times\cite{ray2006,rahaman2010,maurya2015}. For a fairly comprehensive discussion and characterization scheme of spherically charged solutions in general relativity, see Ref.~\cite{ivanov2002}.

Because the electromagnetic stress-energy tensor has a non-zero divergence in the presence of charge, one cannot use it as the sole source in Einstein's field equations: Einstein's tensor has a zero divergence due to the Bianchi identity, and cannot be equated to a tensor with non-zero divergence. Therefore, treating situations with electromagnetic charge in general relativity is even more difficult than in special relativity: without some addition to the electromagnetic stress-energy tensor, one cannot start to solve the simplest problem.

Also, one cannot introduce point particles to supplement the stress-energy tensor: their infinite energy density creates singularities in space-time. Therefore, in the literature where electric charge is studied in general relativity, the electromagnetic stress-energy tensor is augmented typically using a fluid, where the density of the fluid is proportional to the density of its constituent particles, i.e. a continuum approximation. The addition of the fluid results in 6 dynamic degrees of freedom at each point in space-time (3 in the fluid and 3 in the electromagnetic current). This makes the field equations underdetermined (there are only 3 dynamical equations of motion at each point in space-time), and the charge distribution must be set as a model parameter, rather than solved for by the dynamics\cite{ivanov2002}. Interestingly, this makes finding many solutions easier, since one has free parameters to tune\footnote{As Ivanov writes, ``The presence of charge serves as a safety valve, which absorbs much of the fine tuning, necessary in the uncharged case.''\cite{ivanov2002}}.

Recently, some attempts at modeling a ``charged fluid'' appear in the literature, where the electromagnetic charge is stuck on the fluid (to remove the extra degrees of freedom): the energy density of the fluid is tied in an ad-hoc way to the density of the charge. For instance, this has been done (in a spherically static case) by adding a perfect fluid stress-energy tensor to the electromagnetic stress-energy tensor and setting the energy density of the fluid to be proportional to the charge density squared\cite{tiwari1984}. To obtain stable solutions, negative pressure is required (since the charge self-repels), and the equation of state (the relationship between the energy density, $\epsilon$, and pressure, $P$, of the fluid) is set to $P=-\epsilon$\cite{cohen1969,cooperstock1978,lopez1984,gron1985,gautrau1985,ponce1987,tiwari1991,ray2004b,ray2008}. This equation of state has been called the ``false vacuum,'' ``degenerate vacuum'' and ``vacuum fluid'' among other names. All of these attempts center around special cases (e.g. static situations with spherical symmetry), rather than treating the general problem, which again is unsolved.



\section{Mathematical Review of Classical Electrodynamics}

Having reviewed some of the history of the development of electrodynamics, let us now review the current state of the associated mathematics. The electromagnetic field generated by a charge distribution is calculated via Maxwell's equations, and is without pathology. We refer the reader to \cite{jackson} for review. However, as mentioned in the history, due to our lack of knowledge of the bare mass and binding force for fundamental charged objects, a full treatment of the dynamics of a charged object, including self-interaction, is intractable. We now lay down the mathematics of why this is so.

\subsection{\label{sec:notation}Notation}

For the remainder of the paper, the following notation will be used (unless otherwise noted). Relativistic (geometrized) units are used throughout\footnote{Using factors of the speed of light, $c$, and the gravitational constant $G$, all units are converted to units of distance.\cite{misner}}. Capital italicized variables with Greek superscripts or subscripts, e.g. $F_{\mu\nu}$ or $F^{\mu\nu}$, are tensors defined in 4-dimensional space-time; covariant tensors have subscripts and contravariant tensors have superscripts; Greek indices vary from 0 to 3, with the 0$^{\rm th}$ element being the time component, and 1-3 being space components. Lowercase italicized variables with Greek superscripts are tensor densities (a tensor multiplied by $\sqrt{|g|}$, e.g. $j^\mu=\sqrt{|g|}J^\mu$). Bold italic variables, e.g. $\df{F}$, are differential forms (totally antisymmetric covariant tensors). Bold non-italic variables, e.g. $\vec{B}$, are spatial vectors; these may also be represented using Latin indices, e.g. $B^i$; Latin indices vary from 1 to 3. Repeated indices in a product are implicitly summed over.

We assume a space-time characterized by local coordinates $x^\mu=(t,x^i)$, with a metric, $g_{\mu\nu}$ with signature (-+++). The determinant of the metric is written as $g$ (with no indices). The totally antisymmetric (Levi-Civita) tensor is written as $\eta_{\alpha\beta\gamma\delta}=\sqrt{|g|}\epsilon_{\alpha\beta\gamma\delta}$, where $\epsilon_{\alpha\beta\gamma\delta}$ has components $\pm 1,0$. We may also write $\eta^{\alpha\beta\gamma\delta}=\frac{1}{\sqrt{|g|}}\epsilon^{\alpha\beta\gamma\delta}$, where $\epsilon^{\alpha\beta\gamma\delta}=-\epsilon_{\alpha\beta\gamma\delta}$ (more generally, $\epsilon^{\alpha\beta\gamma\delta}=(-1)^{n_e}\epsilon_{\alpha\beta\gamma\delta}$, where $n_e$ is the number of negative eigenvalues of $g_{\mu\nu}$). 

Concerning operators, $\partial_\mu$ or $\partial_i$ is the partial derivative with respect to the coordinate of the subscript, and $\nabla_\mu$ is the covariant derivative using the Levi-Civita connection\cite{misner}. In flat space-time, $\nabla$ is the spatial gradient operator. The exterior derivative\cite{misner,deRham1955}, which operates on differential forms is denoted as $\df{d}$ (taking $n$-forms to $(n+1)$-forms). The Hodge star operator\cite{misner,deRham1955}, which in a $p$ dimensional space, takes an $n$-form to a $(p-n)$-form, is denoted $\df{*}$. The codifferential operator\cite{misner,deRham1955}, $\df{\delta}\equiv\df{*d*}$, takes $n$-forms to $(n-1)$-forms\footnote{For more background on differential forms and exterior calculus, see Refs.\cite{misner,deRham1955}}. 

The covariant representations of kinematic variables are:
\begin{equation}
\begin{array}{rcl}
  r^{\mu}&=&(t,r^i)\\
  v^{\mu}&=&(\gamma,\gamma v^i)
\end{array}
\label{kinVars}
\end{equation}
where $\vec{r}=r^i$ is the position of a point (i.e. the center of mass of an object) at time $t$, both with units of length; $\vec{v}=v^i$ is the unitless fraction of the velocity of a point to the speed of light, $c$ (or equivalently, $c=1$); $\gamma$ is the Lorentz factor, $\gamma=1/\sqrt{1-v^2}$. We use $\vec{p}$ for an object's center-of-mass momentum, and $\vec{\pi}$ to denote a field's momentum density.

The antisymmetric part of a tensor may be written using square brackets in the indices as $A_{[\mu}B_{\nu]}\equiv\frac{1}{2}(A_{\mu}B_{\nu}-B_{\nu}A_{\mu})$; likewise, parentheses in indices represent the symmetric part of a tensor. Square brackets around two operators signifies the commutator, for instance,  $[\nabla_\mu,\nabla_\nu] \equiv \nabla_\mu\nabla_\nu - \nabla_\nu\nabla_\mu$. Square brackets or parentheses elsewhere have no special meaning. 

The covariant representations of the electromagnetic variables are:
\begin{equation}
\begin{array}{rcl}
A^{\mu}&=&(\phi,A^i)\\
F_{\mu \nu}&\equiv&2\nabla_{[\mu}A_{\nu]}~~({\rm or~}\df{F}\equiv\df{d}\df{A})\\
F^{\mu\nu} & = &\left(
\begin{array}{cccc}
0 & E^1 & E^2 & E^3\\
-E^1 & 0 & B^3 & -B^2\\
-E^2 & -B^3 & 0 & B^1\\
-E^3 & B^2 & -B^1 & 0
\end{array}
\right) \\
J^{\mu}&=&(\rho,J^i)\equiv\nabla_\nu F^{\mu\nu},
\end{array}
\label{emVars}
\end{equation}
where $A^\mu$ is the electromagnetic potential (made of the scalar and vector potentials, $\phi$ and $\vec{A}=A^i$), which are unitless; $F^{\mu\nu}$ is the electromagnetic field tensor (made up of the electromagnetic fields, $\vec{E}=E^i$ and $\vec{B}=B^i$), with units of 1/distance; $J^\mu$ is the electromagnetic current density (made up of charge and current density, $\rho$ and $\vec{J}=J^i$), with units of 1/distance$^2$.

The notation $\overset{\int}{=}$ is short-hand for
\begin{equation}
\begin{array}{l}
  b \overset{\int}{=} c \rightarrow  \int b d^4 x = \int c d^4 x,
\label{intEquals}
\end{array}
\end{equation}
where $d^4x=d x^0 d x^1 dx^2 d x^3$, and $b$ and $c$ are scalar densities; the integral is over all space-time, and $b$ and $c$ are assumed to tend toward zero fast enough near infinity, such that integration by parts may be used to remove total divergences.

\subsection{Dynamics of a Charged Object\label{sec:dynamics1}}

In this section, we develop the center-of-mass dynamics of a discrete charged object in an electromagnetic field. First, we will write down local momentum conservation in a general form, and integrate it to arrive at the center-of-mass equations of motion. For simplicity, in this section flat space-time will be assumed, but everything is special-relativistically invariant (although this may not initially be obvious). We make no assumptions on the internal mass model of our charge, only that momentum density is a conserved vector quantity. We also make no assumption on the nature of non-electromagnetic forces; ``force density'' in this section merely means the local increase in momentum density due to that force density (a source term in the continuity equation for momentum). 

Consider a stable distribution of charge, which is bounded in the sense that a surface can be drawn around the distribution, which completely contains the charge. Internally, the distribution has local charge density $\rho$ and current density $\vec{J}$ (which can vary with time and position within the object; we make no constraints on those at present). This electromagnetic current, $J^\mu = (\rho, \vec{J})$, could inherently carry some ``bare'' momentum density in its own right. Call this $\vec{\pi}_{\rm bare}$; this is the momentum density of matter that directly interacts with the electromagnetic field, but note this does {\em not} contain the momentum in the electromagnetic field itself; this would be the momentum of the charge density if it were stripped of its electromagnetic field (in the literature the mass associated with this momentum is called the ``bare mass''\cite{yaghjian}). There is also an associated stress tensor\footnote{This is the spatial part of the stress-energy tensor, which will be discussed more later}, $T_{s,{\rm bare}}$\cite{goldstein1965classical,jackson}; if there is no loss/gain of bare momentum density, it obeys the continuity equation:
\begin{equation}
\frac{\partial}{\partial t}(\vec{\pi}_{\rm bare})+\nabla\cdot T_{s,{\rm bare}}=0.
\label{isolatedMomentumConservation}
\end{equation}
We know the electromagnetic field delivers energy/momentum to charged objects: from Maxwell's equations, in the presence of charge, the electromagnetic field loses momentum density at a rate given by the negative of the Lorentz force density, $\vec{f}_{\rm em}\equiv \rho \vec{E} + \vec{J}\times\vec{B}$\footnote{This makes no assumption on the structure of the charge; this expression of the local loss of momentum from the electromagnetic field can be derived directly from Maxwell's equations assuming $J^\mu$ is defined as $\nabla_\nu F^{\mu\nu}$\cite{jackson}.}. By definition, this must be absorbed by the bare momentum density. It is also possible there is some matter that does not directly interact with the electromagnetic field in our object, but is bound to our bare matter in some way; we'll label this ``other'' momentum density, $\vec{\pi}_{\rm other}$, its associated stress tensor, $T_{s,{\rm other}}$, and the binding force density from the other matter on the bare matter as $\vec{f}_{b, {\rm bare}}$. 

For completeness, allow for some non-electromagnetic, external force density $\vec{f}_{\rm ext, bare}$, which acts from outside objects directly on the bare matter. Then, conservation of momentum density for the bare matter is
\begin{equation}
\rho \vec{E} + \vec{J}\times\vec{B} + \vec{f}_{b, {\rm bare}} + \vec{f}_{\rm ext, bare} = \frac{\partial\vec{\pi}_{\rm bare}}{\partial t}+\nabla\cdot T_{s,{\rm bare}}.
\label{generalMomentumConservation}
\end{equation}

Integrating over a volume that fully contains the charge, such that the integral of the stress tensor over the surface of the volume is zero, gives
\begin{equation}
\int(\rho \vec{E} + \vec{J}\times\vec{B} + \vec{f}_{b, {\rm bare}} + \vec{f}_{\rm ext, bare})dV = \frac{d\vec{p}_{\rm bare}}{dt},
\label{intGeneralMomentumConservation}
\end{equation}
where $dV$ is the spatial volume element, and $\vec{p}_{\rm bare}=\int \vec{\pi}_{\rm bare}dV$ is the integrated (total) bare momentum of the charge (again, not including the contribution from its field).

Conservation of momentum density for the other matter, including an external force, $\vec{f}_{\rm ext, other}$, that acts directly on the other matter, is
\begin{equation}
  \begin{array}{rcl}
    \vec{f}_{b, {\rm other}} + \vec{f}_{\rm ext, other} & = & \frac{\partial\vec{\pi}_{\rm other}}{\partial t}+\nabla\cdot T_{s,{\rm other}},
    \end{array}
\label{otherMomentumConservation}
\end{equation}
where $\vec{f}_{b, {\rm other}}$ is the binding force on the other matter. The integral of Eq.~\ref{otherMomentumConservation} over the object, assuming the other matter is also bounded by our surfaced, is
\begin{equation}
\int(\vec{f}_{b, {\rm other}} + \vec{f}_{\rm ext, other})dV = \frac{d\vec{p}_{\rm other}}{dt},
\label{intOtherMomentumConservation}
\end{equation}
where $\vec{p}_{\rm other}=\int \vec{\pi}_{\rm other}dV$.

Now separate the electromagnetic field into a self-field ($\vec{E}_{\rm self}$ and $\vec{B}_{\rm self}$) due to the object itself, and an external field ($\vec{E}_{\rm ext}$ and $\vec{B}_{\rm ext}$) due to other charges outside the object. Assuming the distribution is sufficiently small compared to the variation of the external electromagnetic field, we can immediately integrate terms with the external field, and total momentum conservation (the sum of Eq.~\ref{intGeneralMomentumConservation} and Eq.~\ref{intOtherMomentumConservation}) becomes:
\begin{equation}
\begin{array}{c}
q \vec{E}_{\rm ext}+q\vec{v}\times\vec{B}_{\rm ext}+\int\left(\rho \vec{E}_{\rm self}+\vec{J}\times\vec{B}_{\rm self}\right)dV + \vec{F}_{\rm other}\\
=\frac{d}{dt}(\vec{p}_{\rm bare} + \vec{p}_{\rm other}),
\end{array}
\label{particleEqOfMotion1}
\end{equation}
where $q=\int\rho dV$ and $\vec{v}=\frac{1}{q}\int\vec{J}dV$; if the charge is sufficiently stable (i.e. rigid), then $\vec{v}$ represents the center-of-mass motion of our compact object.
\begin{equation}
\begin{array}{c}
  \vec{F}_{\rm other}=\int \left(\vec{f}_{\rm ext, bare} + \vec{f}_{\rm ext, other} +\vec{f}_{b, {\rm bare}} + \vec{f}_{b, {\rm other}}\right)dV
\end{array}
\label{fExtIntegral}
\end{equation}
is the total non-electromagnetic force acting on the object: the net momentum added to or removed from the object by non-electromagnetic forces.

The integral of the self-field over the distribution results in the ``field reaction'', i.e. the rate of change of the momentum of the self-electromagnetic field due to the distribution's motion\cite{yaghjian}. The field reaction naturally splits into two pieces: a term that will look like $-\frac{d}{dt}(\gamma m_{\rm field}\vec{v})$, associated with fields bound to the charge, and thus contributing to its inertial mass ($m_{\rm field}$ is that contribution)\cite{yaghjian}; and a term that is associated with the field that eventually escapes the object as radiation, and is hence called the ``radiation reaction''\footnote{Many authors include the contribution to the inertial mass in what they call the ``radiation reaction''\cite{yaghjian}}\cite{yaghjian,galley2010,forgacs2012,galley2012}.

While one can obtain $m_{\rm field}$ by extracting it from the self-field integral of Eq.~\ref{particleEqOfMotion1}, it is easier to use mass-energy equivalence from special relativity. The field energy (and hence $m_{\rm field}$) is given by\cite{jackson}
\begin{equation}
m_{\rm field} = \int\left(\frac{1}{2}E_{\rm self, rest}^2+\frac{1}{2}B_{\rm self, rest}^2\right)dV,
\label{fieldEnergy}
\end{equation}
where the fields are evaluated when the charge is isolated and at rest, and the integral is over all space. This integral is linearly inversely proportional to the size of the object (this is easy to show, for example, for a sphere of charge)\cite{jackson,griffiths,yaghjian}. Therefore, this self-energy, and its associated inertial mass, approaches infinity as one takes the point charge limit, leading to the ``infinite self-energy'' problem\cite{feynman1963}.

The radiation reaction (the recoil felt by the object due to momentum carried away by radiation) again, is what is left of the self-field integral after subtracting off the inertial contribution from $m_{\rm field}$:
\begin{equation}
\begin{array}{l}
({\rm radiation~reaction})\\
=-\int\left(\rho \vec{E}_{\rm self}+\vec{J}\times\vec{B}_{\rm self}\right)dV-\frac{d}{dt}(\gamma m_{\rm field}\vec{v}).
\end{array}
\label{radReactRenorm}
\end{equation}
Note that the radiation reaction stays finite as the size of the charge approaches zero: the part of the self-field integral which approaches infinity can all be rolled into the mass. With the assumption that the integrated momenta are proportional to $\gamma \vec{v}$ (an assumption of sufficient rigidity), we may also now define masses for the different momenta as $\vec{p}_{\rm bare} = \gamma m_{\rm bare}\vec{v}$ and $\vec{p}_{\rm other} = \gamma m_{\rm other}\vec{v}$\footnote{Any required assumption of rigidity will necessarily be violated over short time-scales as changes in external forces propagate across the object; however, this assumption of rigidity is required to produce well-determined equations of motion, as will be discussed in more detail later.}. Replacing the momenta, extracting the contribution to the inertial mass from the self-field integrals, and rearranging Eq.~\ref{particleEqOfMotion1}, we obtain
\begin{equation}
\begin{array}{l}
q \vec{E}_{\rm ext}+q\vec{v}\times\vec{B}_{\rm ext} + \vec{F}_{\rm other}\\
=\frac{d}{dt}(\gamma m\vec{v})+({\rm radiation~reaction}),
\end{array}
\label{particleEqOfMotion2}
\end{equation}
where $m=m_{\rm bare} + m_{\rm field} + m_{\rm other}$ is the total inertial mass (what one would measure as the inertial mass in the laboratory). Assuming that $m_{\rm bare}$ and $m_{\rm other}$ appropriately cancels the infinite $m_{\rm field}$ of a point charge, to give the measured $m$, constitutes ``renormalization''.

Eq.~\ref{particleEqOfMotion2} is the equations of motion for the center-of-mass dynamics of a sufficiently stable charge, and is the force law found in text books for a charged particle, or in papers discussing radiation reaction. We tried to keep this as general as possible: we didn't assume much about the structure of the charge, only that it is small compared to variations of the external field and stable enough that $\vec{v}$ is well-defined (the different momenta are proportional to it). If one ignores the radiation reaction, these equations of motion are readily solvable, and effectively describe many experiments\footnote{The radiation reaction is negligible for many situations and may be ignored without too much effect. See \cite{jackson}, Ch. 16 for a discussion of when radiation reaction becomes important for various experiments.}.

However, including the radiation reaction is much more difficult. The radiation reaction term depends heavily on how charge is distributed in the object. Therefore, in order to solve dynamical problems for the center-of-mass motion of a discrete charged body {\em including the radiation reaction}, one must use a process summarized as follows (this is the process used in all examples in the literature of which the author is aware):

\begin{enumerate}
	\item {Assume an internal distribution of charge for all involved charged bodies (such as rigid spheres, or point charges). This enables solving for the radiation reaction as a function of $\vec{v}$ and its time derivatives. It also implicitly sets the binding force/stress tensor divergences everywhere in the bodies.}
	\item{Assume the mass of each charged body (a measured value if you're treating a real charge), $m$; this is necessary because of the lack of knowledge of how to calculate $m_{\rm bare}$ and $m_{\rm other}$.
	}
	\item{With the radiation reaction known as a function of $\vec{v}$ and its derivatives, Eq.~\ref{particleEqOfMotion2} provides a well-determined system of equations for the center-of-mass dynamics of each body: given initial conditions, Eq.~\ref{particleEqOfMotion2} may be solved.}
\end{enumerate}
As stated in Sec.~\ref{sec:issues}, this process necessarily violates causality on the time scale of light crossing the object.

Also, if individual bodies are too close to each other, our assumption that $\vec{E}_{\rm ext}$, $\vec{B}_{\rm ext}$ are constant over the charge fails. Also, you will not be able to effectively solve for the radiation reaction before solving the dynamics: if the radiation fields significantly overlap, the associated radiated momentum/power does not obey superposition (the fields add, but their momentum/power do not).  Therefore, this methodology is only effective for bodies that do not interact too closely.

While this process may be used to solve for the center-of-mass dynamics of extended, charged bodies under certain circumstances (while unfortunately violating causality on short time scales), solving for the internal dynamics of a charge distribution (which is equivalent to solving systems where charged objects interact closely) is completely intractable without {\em a priori} knowledge of $\pi_{\rm bare}$, $T_{s,{\rm bare}}$, $\pi_{\rm other}$, $T_{s,{\rm other}}$, $\vec{f}_{b, {\rm bare}}$, $\vec{f}_{b, {\rm other}}$, $\vec{f}_{{\rm ext, bare}}$, and $\vec{f}_{{\rm ext, other}}$\footnote{With enough assumptions, one can make some progress in developing internal dynamics without full knowledge of these variables. For instance, one may develop equations of motion by assuming a spherical charge is comprised of spherical shells, which are tied together by some linear restoring force. This gives enough information about the internal binding force, that with some other assumptions on the motion, one can solve for the center-of-mass motion and the induced dipole moment of such a structure; this has been done in \cite{flammer2016}. However, this still violates causality due to requiring that the spherical shell components remain spherical; any predetermination of the form of the charge density of an extended object necessarily violates causality.}.

Without this knowledge, mathematically, the only way to produce well-posed equations of motion, which do not manifestly violate causality, is to take the point charge (particle) limit, where the only motion is the center-of-mass motion. Then, for particles, which do not interact too closely, Eq.~\ref{particleEqOfMotion2} becomes the Lorentz-Abraham-Dirac equations of motion\cite{yaghjian}. Even in the case of closely interacting particles (where the radiated energy/momentum cannot be predetermined), in principle, if one is careful enough, one could track all the electromagnetic momentum/power emitted or absorbed through a small surface surrounding each interacting particle, and use that to calculate the change in momentum of each particle from one time to a slightly later time (assuming you know/measure each particle's mass). Therefore, in the point charge limit, one can create a well-posed, causally correct problem, which can be solved.

Without {\em a priori} knowledge of the form of the internal momentum, stress tensors, and force densities, taking the point-charge limit appears to be the {\em only} way of solving electromagnetic problems in a self-consistent, causally correct way. The cost, mathematically, of taking the point-charge limit, is that $m_{\rm field}$ is infinite, creating the need for renormalization (set $m_{\rm bare}=-\infty$ so $m$ is the measured finite value).

All of these difficulties with developing equations of motion for charged objects may be summarized concisely and covariantly as follows. Conservation of momentum (and energy) density can be written by setting the divergence of a stress-energy tensor (call it $T^{\mu\nu}$) to zero\cite{misner}:
\begin{equation}
\nabla_\mu T^{\mu\nu} = 0.
\label{stressEnergyConservation}
\end{equation}
The electromagnetic stress-energy tensor (the contribution to the stress-energy tensor from the electromagnetic field),
\begin{equation}
  T_{\rm EM}^{\mu \nu}=g_{\alpha \beta}F^{\mu \alpha}F^{\nu \beta}-\frac{1}{4}g^{\mu \nu}F_{\alpha \beta}F^{\alpha \beta},
  \label{stressTens}
\end{equation}
has divergence\cite{misner}
\begin{equation}
\nabla_\mu T_{\rm EM}^{\mu\nu} = J_{\mu}F^{\mu \nu} = (-\vec{J}\cdot\vec{E}, -\rho \vec{E} -\vec{J}\times\vec{B}),
\label{emStressEnergyNonConservation}
\end{equation}
which is manifestly non-zero in the presence of electromagnetic charge. Without some addition to $T^{\mu\nu}$, energy-momentum cannot be conserved. This is the source of all the problems/paradoxes associated with developing classical electrodynamics of extended bodies\cite{tangherlini1963,rohrlich1997,yaghjian,jackson}. Some other contribution to the total stress-energy tensor is necessary to allow the total divergence to be zero, but no reasonable addition has been found, outside of including point charges, with their associated infinite masses\cite{jackson}. If a reasonable non-particle addition were to be included, all of the paradoxes and problems with electromagnetism would be resolved. We need the stress-energy tensor for the bare matter, $T_{\rm bare}^{\mu\nu}$, which can soak up this loss of energy-momentum from the electromagnetic field. If there is other matter, then we also need $T_{\rm other}^{\mu\nu}$ (for other bound matter), $T_{b,{\rm bare}}^{\mu\nu}$ and $T_{b,{\rm other}}^{\mu\nu}$ (for the binding interaction between the bare and other matter); if there is an external force, then we require $T_{\rm ext, bare}^{\mu\nu}$ and $T_{\rm ext, other}^{\mu\nu}$ to describe the energy-momentum transfer for those interactions, and so on.

These would yield the necessary expressions of this section: $\pi_{\rm bare}=T_{\rm bare}^{0 i}$, $T_{s,{\rm bare}}=T_{\rm bare}^{i j}$, $\pi_{\rm other}=T_{\rm other}^{0 i}$, and $T_{s,{\rm other}}=T_{\rm other}^{i j}$. For the interactions, $\vec{f}_{b, {\rm bare}}= -\nabla_\mu T_{b, {\rm bare}}^{\mu i}$, $\vec{f}_{b, {\rm other}} = -\nabla_\mu T_{b,{\rm other}}^{\mu i}$ and $\vec{f}_{{\rm ext, bare}} = -\nabla_\mu T_{\rm ext, bare}^{\mu i}$, $\vec{f}_{{\rm ext, other}} = -\nabla_\mu T_{\rm ext, other}^{\mu i}$, similar to how $\nabla_\mu T_{\rm EM}^{\mu i} = -\vec{f}_{\rm EM}=-\rho \vec{E} -\vec{J}\times\vec{B}$. If we had this knowledge, we could solve the local equations of motion Eq.~\ref{generalMomentumConservation} and Eq.~\ref{otherMomentumConservation} for all of the internal dynamics of an object.


\subsection{\label{sec:leastAction1}Least Action Principle and Deriving Maxwell's Equations}

The principle of least action has a long history, and has been used extensively to develop fundamental theories\cite{goldstein1965classical}. To use the principle of least action to develop a field theory, one defines a scalar ``action integral'', $S$, which is the integral over our entire space-time manifold of a scalar density, the Lagrangian density $\mathcal{L}$, which can depend on various tensor fields, ${{T_i}^{\mu_1, \mu_2...}}_{\nu_1,\nu_2...}$, and their derivatives ($i$ here labels the different fields, not a coordinate), as 
\begin{equation}
  S = \int \mathcal{L}({{T_i}^{\mu_1, \mu_2...}}_{\nu_1,\nu_2...}, \partial_\alpha {{T_i}^{\mu_1, \mu_2...}}_{\nu_1,\nu_2...},...)d^4 x.
  \label{lagrangianDefinition}
\end{equation}
where $d^4x = dx^0 dx^1 dx^2 dx^3$. We've used partial derivatives here rather than covariant derivatives to allow tensor quantities, such as the curvature tensor, which cannot be written as a covariant derivative of its field, but are still tensorial. Note the Lagrangian density must still be a {\em scalar density}, $\mathcal{L} = \sqrt{|g|}L$, where $L$ is a Lorentz scalar, i.e. it is the same in all coordinate systems, so almost always derivatives should be covariant. The derivatives are often, but not always limited to first order\footnote{See \cite{simon1990} for some discussion.}. One then requires variations of this action to be zero against smooth, arbitrary, infinitesimal variations of the different fields (signified by an operator $\delta$)\footnote{Note the variation must obey any inherent symmetries of the underlying field. For example, the variation of the metric must be symmetric since the metric is symmetric by definition.},
\begin{equation}
  \delta S = \int \frac{\partial \mathcal{L}}{\partial {{T_i}^{\mu_1, \mu_2...}}_{\nu_1,\nu_2...}}\delta{{T_i}^{\mu_1, \mu_2...}}_{\nu_1,\nu_2...} d^4 x = 0,
  \label{deltaActionEqualsZero}
\end{equation}
with the understanding that integration by parts is used to convert variations of $\partial_\alpha {T_i^{\mu_1, \mu_2...}}_{\nu_1,\nu_2...}$ to variations of ${T_i^{\mu_1, \mu_2...}}_{\nu_1,\nu_2...}$ and a total divergence, which can be converted to a surface integral and assumed zero (as variations are assumed to be zero on the bounds of integration). Therefore, when we write $\frac{\partial \mathcal{L}}{\partial {{T_i}^{\mu_1, \mu_2...}}_{\nu_1,\nu_2...}}$, what we mean is what multiplies $\delta{{T_i}^{\mu_1, \mu_2...}}_{\nu_1,\nu_2...}$ after integration by parts. Since each field can vary independently, and the variations are allowed to deform in any way in space-time, the following must be true for each of the fields at every point in space-time:
\begin{equation}
  \frac{\partial \mathcal{L}}{\partial {{T_i}^{\mu_1, \mu_2...}}_{\nu_1,\nu_2...}} = 0,
  \label{deltaLEqualsZero}
\end{equation}
where again, the partial derivative with respect to ${T_i^{\mu_1, \mu_2...}}_{\nu_1,\nu_2...}$ is understood to include dependencies on its derivatives via integration by parts. This produces the same number of equations as there are degrees of freedom in all the fields (e.g. 6 equations at each point in space-time for an antisymmetric 2-tensor).

To develop electromagnetism from a Lagrangian density, one starts with (in our units, using the sign convention of Jackson)\cite{jackson}
\begin{equation}
  \mathcal{L}_{\rm EM} = -\frac{1}{4}\sqrt{|g|}F_{\alpha\beta}F^{\alpha\beta}.
  \label{conventionalFreeSpaceLagrangian}
\end{equation}
If the ``field'' is considered to be $A_\alpha$, then
\begin{equation}
  \frac{\partial \mathcal{L}_{\rm EM}}{\partial A_\alpha} = - \sqrt{|g|}\nabla_\beta F^{\alpha\beta},
  \label{conventionalFreeSpaceVariation}
\end{equation}
and the equations of motion for the field are $\nabla_\beta F^{\alpha\beta}=0$, which means there is no electromagnetic charge; Eq.~\ref{conventionalFreeSpaceLagrangian} is the ``free-space Lagrangian''. To allow for charge, an extra term is necessary. To do this, consider an independent vector field $J^\alpha$ (not yet identified as electromagnetic current). Adding the typical ``interaction Lagrangian''\cite{jackson}, $\mathcal{L}_{\rm int}$,  makes the total Lagrangian
\begin{equation}
  \begin{array}{rcl}
    \mathcal{L} & = & \mathcal{L}_{\rm EM} + \mathcal{L}_{\rm int}\\ 
    \mathcal{L}_{\rm int} & = & \sqrt{|g|}J^\alpha A_\alpha.
  \end{array}
  \label{conventionalLagrangian}
\end{equation}
Since $J^\alpha$ is assumed independent of $A_\alpha$, 
\begin{equation}
  \begin{array}{rcl}
    \frac{\partial\mathcal{L}_{\rm int}}{\partial A_\alpha} & = & \sqrt{|g|}J^\alpha \\
    \frac{\partial\mathcal{L}}{\partial A_\alpha} & = & \sqrt{|g|}\left(- \nabla_\beta F^{\alpha\beta} + J^\alpha\right) = 0,
  \end{array}
  \label{conventionalLagrangian2}
\end{equation}
the last equation being the inhomogeneous Maxwell equations with $J^\alpha$ as the electromagnetic current. One can't just vary $J^\alpha$ independent of $A_\alpha$, since its variation would lead to $A^\alpha=0$, which means there would be no field. As before, we unfortunately need to make some assumption about the structure of what is carrying the current. To the author's knowledge, the only option in the literature that yields reasonable equations of motion is the ``particle hypothesis''\cite{milton2006}
\begin{equation}
  J^\alpha \equiv (\rho,\vec{J}) \equiv \sum_i q_i \delta^3(\vec{x}-\vec{r}_i) \left(1, \vec{v}_i\right),
\label{microJDef}
\end{equation}
where $J^\alpha$ is comprised of a flow of particles: $q_i$, $\vec{r}_i$, and  $\vec{v}_i$ are the charge, position, and velocity of the $i^{\rm th}$ point charge (here $i$ is summed over discrete charges, not dimensions), and $\delta^3$ is the 3-dimensional Dirac delta function, not an arbitrary variation.

Using this particle hypothesis, and adding the particle Lagrangian density\cite{goldstein1965classical} to the action, we have the total action:
\begin{equation}
  \begin{array}{rcl}
    S & = & \int \left(\mathcal{L}_{\rm EM} + \mathcal{L}_{\rm int} + \mathcal{L}_{\rm particle}\right) d^4 x \\
    \mathcal{L}_{\rm particle} & = & -\sqrt{|g|} \sum_i \frac{m_i}{\gamma_i} \delta^3(\vec{x}-\vec{r}_i),
  \end{array}
  \label{conventionalParticleAction}
\end{equation}
where $m_i$ is the mass of the $i^{\rm th}$ particle and $\gamma_i$ is its Lorentz factor, $(\sqrt{1-v_i^2})^{-1}$. In the second two terms of the action, we can immediately integrate over the spatial coordinates using the Dirac delta functions,
\begin{equation}
  \begin{array}{rcl}
    S & = & \int \mathcal{L}_{\rm EM} d^4 x \\
    && + \sum_i \int \sqrt{|g|}\left(-q_i \phi + q_i \vec{v}_i\cdot\vec{A} - \frac{m_i}{\gamma_i}\right) dt,
  \end{array}
  \label{conventionalParticleAction2}
\end{equation}
where in the second integral, all fields are implicitly evaluated at $\vec{r}_i$; $\vec{r}_i$ is now varied to arrive at the equations of motion for each particle. Dealing with point charges in curved-space time, using general relativity, is problematic due to the fact that point charges produce space-time singularities, so we will proceed in flat space-time, as is typical. The first integral does not depend on $\vec{r}_i$, and the variation of the second with respect to $\vec{r}_i$ (with the flat-space $g_{\mu\nu}$) yields the Lorentz force equation for each particle\cite{goldstein1965classical,jackson}:
\begin{equation}
  \begin{array}{rcl}
    \frac{d}{dt}\left(\gamma_i m_i \vec{v}_i\right) & = & q \vec{E} +q \vec{v}_i\times\vec{B}.
  \end{array}
  \label{particleLorentzForce}
\end{equation}
However, these are not the correct equations of motion: they are missing the radiation reaction. It isn't surprising that the radiation reaction is left out. It is well known that the principle of least action depends on the system being conservative. In 1900, when Joseph Larmor used the principle of least action to obtain both Maxwell's equations and the Lorentz force\cite{larmor1900}, at the beginning of his treatment, he states
\begin{quote} 
If the individual molecules are to be permanent, the system...must be conservative; so that the Principle of Least Action supplies a foundation certainly wide enough...
\end{quote}
With the understanding that charged particles inherently can lose a significant amount of energy due to radiation, this argument doesn't hold for a charged particle in general, and calls into question the ability to use the principle of least action for electromagnetism with point charges\footnote{There have been efforts to contrive a Lagrangian for charged objects, which directly includes radiation reaction. For instance, researchers have developed Lagrangians for such dissipative systems by adding in time-reversed copies, doubling the phase space, but producing something where energy is conserved\cite{barone2007}. In any case, one does not obtain anything like the Lagrangians used in the standard model, and such Lagrangians are dependent on the geometry of the charge, so we'd be back to pre-defining the geometry and, for extended objects, violating causality.}.

This lack of self-interaction/radiation in particle theories, which are developed using the principle of least action, is apparent in both classical and quantum mechanics. In quantum electrodynamics, radiative/self-interaction effects are absent until they are added in (after the fact) via perturbation theory, using the construct of virtual particles and virtual photons\footnote{The interaction with virtual particles is indistinguishable from self-interaction. See Sec.~3 from \cite{feynman1949}.}. All of this is to say that the principle of least action does not appear to be able to help us develop equations of motion for point charges in a non-perturbative way. 

However, for non-point charges, the situation is less grim. In fact, self-interaction for extended charged objects, where the charge/current density is bounded, does not need to be accounted for at all at the fundamental level. If the current density, $J^\mu$, is bounded, then in a small region with volume $dV$, the magnitude of the self-field is proportional to $dV$, and the charge enclosed is proportional to $dV$, so the self-force is proportional to $dV^2$. Whereas the force from a finite external field (the field generated by charge/current outside of $dV$) is proportional to $dV$. Therefore, in the limit of $dV\rightarrow 0$, the self-interaction is negligible compared to the interaction with the external field.

That is why with bounded $J^\mu$, conservation of momentum locally is given by Eq.~\ref{generalMomentumConservation}, without any explicit representation of the self-interaction. The local equations of motion are conservative; it is not until an integral is performed over a finite charge that the radiation reaction appears, as in Eq.~\ref{particleEqOfMotion2}. So for extended objects, we could use the principle of least action in a non-perturbative way.

What can we do other than Eq.~\ref{conventionalLagrangian}? It is commonly understood that the Lagrangian of Eq.~\ref{conventionalLagrangian} is required to reproduce the inhomogeneous Maxwell's equations\cite{jackson}. But there is some subtlety here. Recall $J^\mu$ is an independent field from $A^\mu$ in Eq.~\ref{conventionalLagrangian} (they vary separately when varying the Lagrangian). While as far as the author can tell, $\mathcal{L}_{\rm int}$ is the only addition that will assign an independent field as the source in the inhomogeneous Maxwell's equations, that's an assumption we don't have to make. If we don't assume that an independent field, such as a flow of point charges, is the source of electromagnetism, we can still reproduce all of Maxwell's equations.

With the definition (written as a tensor equation or in the language of differential forms)
\begin{equation}
  \begin{array}{rcl}
    F_{\mu\nu} & \equiv & 2 \nabla_{[\mu}A_{\nu]} \\
    \df{F} & \equiv & \df{dA},
  \end{array}
\label{emFieldDef}
\end{equation}
the homogeneous Maxwell's equations
\begin{equation}
  \begin{array}{c}
    \nabla_\alpha F_{\beta\gamma}+\nabla_\beta F_{\gamma\alpha}+\nabla_\gamma F_{\alpha\beta}=0 \\
    \df{dF}=\df{ddA}=0 
  \end{array}
\end{equation}
are identically true (since $\df{dd}$ of any form is zero). This is simply an identity and not the result of any Lagrangian variation\cite{misner}; the {\em inhomogeneous} Maxwell's equations are what $L_{\rm int}$ was meant to produce, since without it, we get $\nabla_\nu F^{\mu\nu}=0$, which means there is no charge. Some addition that is non-zero inside of a charged object is necessary. But the author points out that any suitable addition to the Lagrangian that is non-zero inside of a charged object, but zero outside, may function just as well as $\mathcal{L}_{\rm int}$, as long as its variation with respect to $A^\mu$ can cancel $\nabla_\nu F^{\mu\nu}$, where charge is present. Concerning what would be measured as the electromagnetic current in a laboratory, the current {\em defined} as
\begin{equation}
  \begin{array}{rcl}
    J_\mu & \equiv & \nabla^\nu F_{\mu\nu} \\
    {\df J} & \equiv & \df{*d*F}
  \end{array}
  \label{fluidJDef}
\end{equation}
is guaranteed to be conserved identically, from the antisymmetry of $F_{\mu\nu}$\cite{misner}:
\begin{equation}
  \begin{array}{rcl}
    \nabla_\mu J^\mu & = & \nabla_\mu \nabla_\nu F^{\mu\nu} = 0 \\
    \df{-*d*J} & = & -\df{*dd*F} = 0.
  \end{array}
  \label{fluidJConservation}
\end{equation}
No matter what the internal dynamics of a charged ``particle'' is, the integral of $J^\mu\equiv\nabla_\nu F^{\mu\nu}$ over a sufficiently small particle will result in a conserved 4-current that is indistinguishable from the point-charge current of Eq.~\ref{microJDef}, and is the source of the particle's electromagnetic field. This is all that has been measured experimentally, so from an experimental standpoint, the inhomogeneous Maxwell's equations will be unchanged by any choice of internal Lagrangian for a charged object: we have complete freedom to explore our options.

To summarize, without making any assumptions or constraints on $A^\mu$, and using the two definitions Eqs.~\ref{emFieldDef}, \ref{fluidJDef}, we have all of Maxwell's equations; we obtain the inhomogeneous Maxwell's equations by simply defining it to be true (as long as we don't define $J^\mu$ in any other way, or contradict it, we're free to define it as in Eq.~\ref{fluidJDef}). Or if one likes, we never introduce $J^\mu$ at all; in any case, it's just short-hand for $\nabla_\nu F^{\mu\nu}$. Of course, this isn't anything new. In almost any introductory electromagnetism text\cite{jackson}, or general relativity text for that matter\cite{misner}, electromagnetic current is defined by Eq.~\ref{fluidJDef}, but from the perspective of developing electromagnetism from a Lagrangian, the literature appears to incorrectly imply $\mathcal{L}_{\rm int}$ is required to reproduce the inhomogeneous Maxwell's equations; this is only the case to identify a field that is {\em independent} of $A_\mu$ as electromagnetic current.  


For the remainder of the paper, we will remove $\mathcal{L}_{\rm int}$ and $\mathcal{L}_{\rm particle}$ from the Lagrangian and their associated stress-energy tensor. We use as a starting point only $\mathcal{L}_{\rm EM}$ and its associated stress-energy tensor, $T_{\rm EM}^{\mu\nu}$.

\section{Completing the Stress-Energy Tensor}

We'll return to the principle of least action later. Since the non-zero divergence of the electromagnetic stress-energy tensor is central to the inconsistencies plaguing electromagnetic theory of extended bodies, let us take some time to investigate possible ways to ``complete'' the electromagnetic stress-energy tensor directly, by finding additions that can produce energy-momentum conservation equations that are self-consistent and solvable. To summarize our problem from an energetic standpoint, electrodynamic theory suffers from the fact that the electromagnetic stress-energy tensor, Eq.~\ref{stressTens}, in the presence of electric charge, has manifestly non-zero divergence; from a purely mathematical perspective, one obtains unsolvable problems without an appropriate addition. The only suitable addition that has been found is that of charged point particles. This point-charge assumption produces solvable equations in flat space-time, but at a cost of infinite self-energies. In flat space-time, these may be dealt with via a renormalization procedure, but unfortunately preclude integrating with general relativity. This method also requires perturbative methods to include remaining non-infinite self-interaction terms (radiative corrections).

In this section, we ask the question: is there a purely electromagnetic solution? More explicitly, there are 3 degrees of freedom per point in space-time in the electromagnetic field: 4 in $A^\alpha$ minus 1 since you have one degree of gauge freedom (or if you like, 4 in $J^\alpha$ minus 1 since it is identically conserved). There are 4 energy-momentum conservation equations, but often energy conservation can be related to momentum conservation, which would yield 3 equations of motion per point in space-time. Without adding any more degrees of freedom per point in space-time, what can we add to $T_{\rm EM}^{\mu\nu}$ to produce self-consistent equations of motion\footnote{In the language of Sec.~\ref{sec:dynamics1}, this amounts to setting $T_{\rm other}^{\mu\nu}=T_b^{\mu\nu}=T_{\rm ext}^{\mu\nu}=0$, and finding possible expressions for $T_{\rm bare}^{\mu\nu}$.}?

\subsection{Integrability\label{sec:integrability}}

Since we only have 3 degrees of freedom per point in space-time in the electromagnetic field, we need to ensure that the energy-momentum conservation equations also only result in 3 independent equations of motion, so as not to overdetermine its evolution. This hopefully will allow us to formulate a dynamic system of equations that, given initial conditions, can be solved.

If our resulting conservation of energy-momentum satisfies a scalar identity, leaving us with only 3 independent equations of motion, which with conservation of charge, uniquely evolves the electromagnetic current, that would suffice. The electromagnetic field energy-momentum provides us with just such an identity. As mentioned before, $\nabla_\mu T_{\rm EM}^{\mu \nu}$ is: 
\begin{equation}
\nabla_{\mu}T_{\rm EM}^{\mu \nu}=J_{\mu}F^{\mu \nu}.
\label{divStressTens}
\end{equation}
While this divergence is non-zero, it has the property that it is identically orthogonal to $J_\nu$:
\begin{equation}
J_\nu \nabla_{\mu}T_{\rm EM}^{\mu \nu}=J_\nu J_{\mu} F^{\mu \nu} = 0,
\label{emStressOrthogonalToJ}
\end{equation}
since $J_\nu J_{\mu}$ is symmetric and $F^{\mu \nu}$ is anti-symmetric. This can be interpreted physically in the following way: in a local frame where $\vec{J}=0$, the power delivered from/to the electromagnetic field ($\nabla_{\mu}T_{\rm EM}^{\mu 0}$) is identically 0. This is obvious from the expression $J_{\mu} F^{\mu 0}=-\vec{J}\cdot\vec{E}$.

This is analogous to a similar condition for particles. For a particle with 4-momentum, $p^\mu$, the rate of change of $p^\mu$ in the direction of a time-like vector, $t^\nu$, that is parallel to the particle's path, is $t^\nu\nabla_\nu p^\mu$. This rate of change and $p^\mu$ are identically perpendicular, if the mass of the particle is constant,
\begin{equation}
p_\mu t^\nu \nabla_\nu p^\mu=\frac{1}{2}t^\nu\nabla_\nu\left(p^\mu p_\mu\right)=\frac{1}{2}t^\nu\nabla_\nu\left(-m^2\right)=0.
\label{particleForceOrthogonalToJ}
\end{equation}
This can be seen as being from the fact that in a frame where $p^i=0$ (the instantaneous center-of-mass frame), the energy is a minimum compared to frames boosted out of the center-of-mass frame (a non-moving particle has less energy than a moving one). For an accelerated particle as it passes through its center-of-mass frame, the power delivered must be zero due to this minimum.

Whatever 4-force is applied to particles must have the same property to produce consistent energy and momentum evolution equations: it must be identically perpendicular to $p^\mu$ (this is equivalent to requiring that the power delivered by a force, $\vec{F}$, on a particle is $\vec{F}\cdot\vec{v}$, where $\vec{v}$ is the velocity of the particle). For the case of charged particles, the electromagnetic current, $J^\mu$, and the momentum, $p^\mu$, coincide up to a constant, and the two identities, Eqs.~\ref{emStressOrthogonalToJ} and \ref{particleForceOrthogonalToJ}, are consistent.

In our search for some non-particle matter to absorb the energy-momentum lost by the electromagnetic field in the presence of $J^\mu$, let us use this same identity, and see if there are any appropriate additions that can identically satisfy Eq.~\ref{emStressOrthogonalToJ}. Qualitatively, this means the energy density contained in our addition must be at an extremum in frames where $\vec{J}=0$, such that power density delivered in that frame will be 0.

Mathematically, if we write the complete stress-energy tensor $T^{\mu \nu}$, comprised of $T_{\rm EM}^{\mu \nu}$ and some addition, $T_{\rm add}^{\mu \nu}$,
\begin{equation}
T^{\mu\nu}\equiv T_{\rm EM}^{\mu \nu} + T_{\rm add}^{\mu \nu},
\label{stressEnergyDef}
\end{equation}
we seek additions such that:
\begin{equation}
J_\nu \nabla_\mu T_{\rm add}^{\mu \nu} = 0.
\label{addStressOrthogonalToJ}
\end{equation}
This will ensure energy density conservation is automatically conserved if momentum density is conserved (just as for particles the power is determined by the force as $\vec{F}\cdot\vec{v}$), leaving us with 3 independent equations to determine our 3 degrees of freedom\footnote{For space-like currents, this identity can make one of the momentum equations redundant rather than the energy equation as we will see shortly. In any case, there are 3 independent equations.}.

\subsection{\label{addition}One Possible Addition to the Stress-Energy Tensor}
Using the guidance from the last section, we will require the divergence of $T_{\rm add}^{\mu\nu}$ to be orthogonal to $J^\mu$ identically, and the most obvious choices are terms that explicitly include $J^\mu$. There are only two quadratic, symmetric 2-tensors, which involve the current:
\begin{equation}
J^{\mu}J^{\nu},~~~~g^{\mu \nu}J_{\alpha}J^{\alpha},
\label{possibilities}
\end{equation}
which suggests the following addition,
\begin{equation}
T_{\rm add}^{\mu \nu}=ag^{\mu \nu}J_{\alpha}J^{\alpha}+bJ^{\mu}J^{\nu}.
\label{tMuNuAddGuess}
\end{equation}
where $a$ and $b$ are constants. Taking the divergence of $T_{\rm add}^{\mu \nu}$ yields
\begin{equation}
\begin{array}{rcl}
\nabla_\mu T_{\rm add}^{\mu \nu} & = & ag^{\mu \nu}\nabla_\mu(J_{\alpha}J^{\alpha})+b\nabla_\mu(J^{\mu}J^{\nu}) \\
 & = & 2 a g^{\mu \nu} J_{\alpha}\nabla_\mu J^{\alpha} + b(\nabla_\mu J^{\mu}J^{\nu} + J^{\mu}\nabla_\mu J^{\nu}) \\
 & = & 2 a J_{\mu}\nabla^\nu J^{\mu} + b J_{\mu}\nabla^\mu J^{\nu},
\end{array}
\label{tMuNuAddDiv}
\end{equation}
and taking $a=-\frac{1}{2}b$, we have 
\begin{equation}
\begin{array}{rcl}
\nabla_\mu T_{\rm add}^{\mu \nu} & = & b J_{\mu}\left(\nabla^\mu J^{\nu} -  \nabla^\nu J^{\mu}\right),
\end{array}
\label{tMuNuAdd1}
\end{equation}
which is perpendicular to $J_\nu$, and Eq.~\ref{tMuNuAdd1} satisfies Eq.~\ref{addStressOrthogonalToJ} (the energy density of $T_{\rm add}^{\mu \nu}$ is an extremum in a frame where $\vec{J}$=0). Therefore, one possible addition to the electromagnetic stress-energy tensor is

\begin{equation}
T_{J}^{\mu \nu}\equiv k_J\left(J^{\mu}J^{\nu}-\frac{1}{2}g^{\mu \nu}J_{\alpha}J^{\alpha}\right),
\label{tMuNuAdd}
\end{equation}
where $k_J$ is a constant with units of distance squared. The form of $T_{J}^{\mu \nu}$ is similar to $T_{\rm EM}^{\mu \nu}$. The components explicitly in flat space-time are
\begin{equation}
\begin{array}{rcl}
T_{J}^{00}&=&\frac{k_J}{2}\left(\rho^2+J^2\right)\\
T_{J}^{0k}&=&k_J\rho J^k\\
T_{J}^{ik}&=&k_JJ^{i}J^{k}+\frac{k_J}{2}g^{ik}(\rho^2-J^2),
\end{array}
\label{tMuNuComp}
\end{equation}
where $g^{ik}$ in flat space-time is the Kronecker delta function. The $00$ component looks like what one might guess for the energy stored in a charge density (it has something that looks like a rest term, $\rho^2$, and a kinetic term, $J^2$; with $k_J>0$, the energy density is a minimum in the frame where $\vec{J}=0$). The $0k$ components also look like what one might guess for the momentum carried by a current.

Taking the divergence of $T^{\mu\nu}$ yields the equations of motion:
\begin{equation}
\begin{array}{rcl}
\nabla^{\mu}T_{\mu \nu}&=&\nabla^{\mu}T_{{\rm EM},\mu \nu}+\nabla^{\mu}T_{{J},\mu \nu}\\
& = & J^\mu\left(F_{\mu\nu}+2 k_J \partial_{[\mu}J_{\nu]}\right) = 0.
\end{array}
\label{genEqsOfMotion}
\end{equation}
Note that we have replaced some covariant derivatives with partial derivatives, since (in the absence of torsion), the antisymmetric derivatives coincide.

Let's explore them in flat space-time, since it will reveal some interesting properties of our new stress-energy tensor. Taking into account charge conservation, they are:
\begin{equation}
\begin{array}{l}
\nabla_{\mu}T^{\mu \nu}= 0 = \\
\left(
\begin{array}{c}
-\vec{J}\cdot\vec{E}+k_J\left(\vec{J}\cdot\frac{\partial \vec{J}}{\partial t}+\vec{J}\cdot\nabla\rho\right)\\
-(\rho \vec{E}+\vec{J}\times\vec{B})+k_J\left(\rho\frac{\partial\vec{J}}{\partial t}+\rho\nabla\rho-\vec{J}\times(\nabla\times\vec{J})\right)
\end{array}
\right).
\end{array}
\label{eqsOfMotion}
\end{equation}
The time component (power equation) is redundant by Eq.~\ref{addStressOrthogonalToJ} as long a $\rho$ is non-zero (dot the momentum equation with $\vec{J}$ and divide that by $\rho$), which is guaranteed for time-like currents\footnote{For space-like currents where $\rho$ is zero but $\vec{J}$ is not, the power equation provides one equation, and the momentum equation is guaranteed to be perpendicular to $\vec{J}$ leaving only two independent equations there.}. Thus, for time-like currents all the information contained in Eq.~\ref{eqsOfMotion} may be written as
\begin{equation}
\rho \vec{E}+\vec{J}\times\vec{B}=k_J\left(\rho\frac{\partial\vec{J}}{\partial t}+\rho\nabla\rho-\vec{J}\times(\nabla\times\vec{J})\right).
\label{eqOfMotion}
\end{equation}
This is a well-defined, local force law on the current density, which given initial conditions, can be evolved. Had we chosen any other relationship between $a$ and $b$, we would have found inconsistent momentum and energy equations.

Eq.~\ref{eqOfMotion} is reminiscent of the equations of motion of a fluid. Using the identity $\vec{J}\times(\nabla\times\vec{J})=\frac{1}{2}\nabla(J^2)-(\vec{J}\cdot\nabla)\vec{J}$, and with some algebra, the force law becomes
\begin{equation}
k_J\rho \frac{\partial \vec{J}}{\partial t}+k_J(\vec{J}\cdot\nabla)\vec{J}=-\nabla\left(\frac{k_J}{2}\left(\rho^2-J^2\right)\right) +\rho \vec{E}+\vec{J}\times\vec{B}.
\label{fluid1}
\end{equation}
This is a Navier-Stokes-like equation, where the left hand side represents the total change in momentum of the fluid. The right hand side has a pressure-like term, with pressure $P=\frac{k_J}{2}(\rho^2-J^2)$, and a body force from the electromagnetic field.

Eq.~\ref{fluid1} can be made to look exactly like the Navier-Stokes equation by making the replacement $\vec{J}=\rho \vec{u}$ (again, if $\rho$ is non-zero):
\begin{equation}
  \begin{array}{l}
    k_J\rho^2\left(\frac{\partial \vec{u}}{\partial t}+(\vec{u}\cdot\nabla)\vec{u}\right) \\
    =-\nabla P + k_J \rho^2(\nabla\cdot\vec{u})\vec{u} +\rho \vec{E}+\vec{J}\times\vec{B}.
    \end{array}
\label{fluid2}
\end{equation}
This is now the Navier-Stokes equation\cite{navier1822} with no viscosity for a fluid of mass density $k_J \rho^2$, velocity $\vec{u}$, pressure $P=\frac{k_J}{2}(\rho^2-J^2)$, and a body force, $\rho^2(\nabla\cdot\vec{u})\vec{u} +\rho \vec{E}+\vec{J}\times\vec{B}$\footnote{Note the ``mass'' conservation law for this fluid is slightly different than for a typical fluid. Using conservation of charge, one finds $\frac{\partial(\rho^2)}{\partial t}+2\rho^2\nabla\cdot\vec{u}+\vec{u}\cdot\nabla(\rho^2)=0$; the factor of 2 on the second term is not found in the typical conservation of mass equation associated with the Navier-Stokes equation.}.

Covariantly, a perfect (non-viscous) fluid has a stress-energy tensor given by
\begin{equation}
T_{\rm pf}^{\mu\nu}=(\epsilon+P)u^\mu u^\nu+Pg^{\mu\nu},
\end{equation}
where $\epsilon$ is the energy density, $P$ is the pressure, and $u^\mu$ is the fluid 4-velocity, which satisfies $u_\mu u^\mu = -1$\cite{tolman1939,jackson}. Writing the electromagnetic 4-current as $J^\mu=\sqrt{-J_\alpha J^\alpha}u^\mu=\sqrt{\rho^2-J^2}u^\mu$ (assuming time-like currents), $T_{J}^{\mu\nu}$ takes on the form of the stress-energy tensor of a perfect fluid with the following equation of state, 
\begin{equation}
\epsilon=P=-\frac{k_J}{2}J_\mu J^\mu=\frac{k_J}{2}\left(\rho^2-J^2\right).
\label{addEqOfState}
\end{equation}

To review, we've established conservation of momentum equations that can, for the first time in a self-consistent way to the author's knowledge, describe the evolution of an extended charged object. In this theory, $k_J$ has the role of a fundamental physical constant, like the gravitational constant, $G$, or the speed of light, $c$. Given general relativity and electromagnetism, we could always reduce variables to some unit of distance (using appropriate factors of $G$ and $c$), like in the geometric units we use in this paper. In quantum mechanics, we can do away with all units using $\hbar$ (or equivalently the Planck length) to set a fundamental length scale. Similarly, in the equations of motion, Eq.~\ref{genEqsOfMotion}, $k_J$ sets the length scale for any dynamical problem. One could rewrite all the equations in completely dimensionless form by modifying the $\nabla_\mu$ operator to be unitless using $k_J$, and adding appropriate powers of $k_J$ to all variables to also make them unitless.

Because $T_{J}^{\mu\nu}$ results in equations of motion so similar to the Navier-Stokes equation, this may allow use of well established methods to solve the equations of motion (for instance, to search for stable solutions). The connection to a relativistic perfect fluid should also make available various existing methods for solving these equations in the context of general relativity.

While it may be self-consistent, it may not be representative of the physical world. We will inspect the properties and solutions of this theory later. For now, let us see how we can arrive at this theory using variational methods, which will be informative on how other extended body theories may be developed.

\subsection{\label{sec:leastActionRevisited}Principle of Least Action Revisited}

Having found a self-consistent stress-energy tensor. We turn our attention to whether a suitable Lagrangian exists that will generate the stress-energy tensor of Eq.~\ref{tMuNuAdd}. In general relativity, one can associate a Lagrangian with its contribution to the stress-energy tensor as\cite{carroll1997}
\begin{equation}
\begin{array}{rcl}
\frac{\partial\mathcal{L}}{\partial g_{\mu\nu}} & = & \frac{1}{2}\sqrt{|g|}T^{\mu\nu}.
\end{array}
\label{LTRelation}
\end{equation}

The connection of our addition to a perfect fluid makes this possible, in the case of time-like currents. In \cite{minazzoli2012}, the Lagrangian density is derived for a barotropic fluid (a fluid whose pressure/energy are only functions of the rest mass density). They show that given a conservation law (conservation of matter in \cite{minazzoli2012}) $\nabla_\mu(\rho_m u^\mu)=0$, where $u^\mu$ is the time-like 4-velocity of the fluid and $\rho_m$ is the rest mass density, one can relate the variation of $\rho_m$ to the variation of the metric as\cite{minazzoli2012,harko2010}
\begin{equation}
\delta \rho_m = \frac{1}{2}(g_{\mu\nu}+u_\mu u_\nu) \delta g^{\mu\nu}.
\label{deltaRhoM}
\end{equation}
As in Sec.~\ref{addition}, we cast the electromagnetic current as $J^\mu=\sqrt{-J_\alpha J^\alpha}u^\mu$, and conservation of charge takes the form $\nabla_\mu(\rho_m u^\mu)=0$ with $\rho_m=\sqrt{-J_\alpha J^\alpha}$. Our pressure and energy are then $P=\epsilon=\frac{1}{2}k_J \rho_m^2$. In \cite{minazzoli2012}, they show that if the pressure can be written purely as a function of $\rho_m$, the Lagrangian density that produces the perfect fluid stress-energy tensor is $-\sqrt{|g|}\epsilon$, and the energy density must obey
\begin{equation}
\epsilon = C \rho_m+\rho_m\int_0^{\rho_m} \frac{P(\rho_m')}{\rho_m'^2} d\rho_m',
\label{energyFromFluidLagrangian}
\end{equation}
where $C$ is an arbitrary integration constant. In our case, $P$ and $\epsilon$ are pure functions of $\rho_m$, and if we set $C=0$, our energy is indeed given by Eq.~\ref{energyFromFluidLagrangian}. Since we satisfy all the requirements of \cite{minazzoli2012}, we can say, for time-like currents the Lagrangian density, which produces $T_{J}^{\mu\nu}$, is
\begin{equation}
  \begin{array}{c}
    k_J \mathcal{L}_{J}\\
    \mathcal{L}_J \equiv \frac{1}{2}\sqrt{|g|} J_\alpha J^\alpha.
    \end{array}
\label{addLagrangian}
\end{equation}

However, note there is an implicit assumption here of what is held constant during the metric variation to arrive at the result of \cite{minazzoli2012}; charge conservation was explicitly maintained during metric variation. How can we do that more directly in terms of electromagnetic variables? The simplest way is to to hold the current density,
\begin{equation}
j^{\mu} \equiv \sqrt{|g|} J^\mu,
\label{jDensDef}
\end{equation}
constant during metric variation. This is because charge conservation,
\begin{equation}
\nabla_\mu J^\mu = \frac{1}{\sqrt{|g|}}\partial_\mu\left(\sqrt{|g|}J^\mu\right) = 0,
\label{jDensityCons}
\end{equation}
can be written as $\partial_\mu j^\mu=0$, completely independent of the metric (even in curved space-time). If $j^\mu$ is held constant during metric variation, then charge is guaranteed to be conserved independent of the metric\footnote{Note electromagnetic charge is guaranteed to be conserved independent of the metric by its definition; this is to say if $J^\mu$ were not identically conserved, holding $j^\mu$ constant during metric variation would guarantee its conservation.}.

It is simple to write $\mathcal{L}_{J}$ as a function of $j^\mu$ and the metric:
\begin{equation}
\mathcal{L}_{J} = \frac{1}{2\sqrt{|g|}} g_{\mu\alpha}j^\mu j^\alpha.
\label{addLagrangian2}
\end{equation}
Varying $g_{\mu\nu}$ holding $j^\mu$ constant gives\footnote{$\delta\sqrt{|g|}=\frac{1}{2}\sqrt{|g|}g^{\mu\nu}\delta g_{\mu\nu}$, see \cite{misner}}
\begin{equation}
\begin{array}{rcl}
k_J\left(\frac{\partial\mathcal{L}_{J}}{\partial g_{\mu\nu}}\right)_{j^\mu={\rm const}} & = & \frac{1}{2}\sqrt{|g|}k_J \left(J^\mu J^\nu -\frac{1}{2}g^{\mu\nu}J_\alpha J^\alpha\right) \\
& = & \frac{1}{2}\sqrt{|g|}T_{J}^{\mu\nu},
\end{array}
\label{addLagrangianVariation}
\end{equation}
which is the correct relationship between a Lagrangian and its stress-energy tensor. Therefore, using charge conservation as an argument to hold $j^\mu$ constant, we obtain the same result as \cite{minazzoli2012} (but without any time-like current assumption): our Lagrangian density is given by Eq.~\ref{addLagrangian} (or Eq.~\ref{addLagrangian2}), which produces the correct stress-energy tensor.

This Lagrangian has been studied before\cite{bopp1940,podolsky1942,zayats2014,gratus2015} in what has come to be called ``Bopp-Podolsky'' electrodynamics. However, in all the cases the author is aware of, these studies maintain $\mathcal{L}_{\rm int}$ and $\mathcal{L}_{\rm particle}$ in the Lagrangian, and are interested in how these terms modify the point-charge theory. We have done something different; we have removed point charges from the theory altogether, to obtain a point-charge-free theory that potentially could yield small, stable, charged (but non-point-charge) solutions.  

It's also worth noting that holding $j^\mu$ constant is a departure from what is typically done in developing electromagnetism from a Lagrangian, as in Sec.~\ref{sec:leastAction1} or Bopp-Podolsky electrodynamics, where $A_\mu$ is held constant during metric variation, and then independently varied in its own right to obtain Maxwell's inhomogeneous equations. What are the consequences of using $j^\mu$ rather than $A_\mu$ as independent from the metric? Possibly most importantly, what does this do to the electromagnetic stress-energy tensor? If that has changed form, our original goal of trying find an addition that is consistent with it is moot.

To address this question, it's useful to introduce the electromagnetic field tensor density,
\begin{equation}
f^{\mu\nu} \equiv \sqrt{|g|} F^{\mu\nu},
\label{fDensDef}
\end{equation}
and using the relation $j^\mu = \partial_\nu f^{\mu\nu}$ (even in curved space-time), the variation of $f^{\mu\nu}$ and $j^\mu$ are simply related by 
\begin{equation}
\begin{array}{l}
  B_\mu \delta j^\mu \overset{\int}{=} \nabla_{[\mu} B_{\nu]}\delta f^{\mu\nu},
\label{deltaJToDeltaF}
\end{array}
\end{equation}
for any one-form $B_\mu$ (see Sec.~\ref{sec:notation} for the definition of $\overset{\int}{=}$).

Eq.~\ref{deltaJToDeltaF} does not involve the metric at all (treating $f^{\mu\nu}$ or $j^\mu$ as independent results in the same Einstein's equations with the same stress-energy tensor). Now let us address the electromagnetic field Lagrangian:
\begin{equation}
\begin{array}{rcl}
\mathcal{L}_{\rm EM} & = & -\frac{1}{4}\sqrt{|g|}F_{\alpha\beta}F^{\alpha\beta}\\
 & = & -\frac{1}{4\sqrt{|g|}}g_{\mu\alpha}g_{\nu\beta}f^{\mu\nu}f^{\alpha\beta}.
\end{array}
\label{emFieldLagrangian}
\end{equation}
The conventional variation, considering $A_\mu$ as the independent EM field, is\cite{jackson}
\begin{equation}
\begin{array}{rcl}
\delta\mathcal{L}_{\rm EM} & \overset{\int}{=} & \frac{1}{2}\sqrt{|g|}T_{{\rm EM}}^{\mu\nu}\delta g_{\mu\nu}- j^\mu \delta A_\mu.
\end{array}
\label{emFieldVariationOld}
\end{equation}
Calculating the variation using $f^{\mu\nu}$ as independent results in
\begin{equation}
\begin{array}{rcl}
\delta\mathcal{L}_{\rm EM} & = & -\frac{1}{2}\sqrt{|g|}T_{{\rm EM}}^{\mu\nu}\delta g_{\mu\nu}-\frac{1}{2}F_{\mu\nu} \delta f^{\mu\nu}.
\end{array}
\label{totEMVariation}
\end{equation}
From this, it's clear that using $j^\mu$ as independent results in
\begin{equation}
\begin{array}{rcl}
\delta\mathcal{L}_{\rm EM} & \overset{\int}{=} & -\frac{1}{2}\sqrt{|g|}T_{{\rm EM}}^{\mu\nu}\delta g_{\mu\nu}-A_\mu \delta j^\mu.
\end{array}
\label{variationEMLagrangianCharge}
\end{equation}
Whether we use $j^\mu$ or $A_\mu$ as independent, we still get the same, standard electromagnetic stress-energy tensor from $\mathcal{L}_{\rm EM}$, although with opposite sign. The sign is of little importance, however, since we can change its sign in the total Lagrangian, or other signs, such as of $k_J$, to compensate as necessary (the overall sign of the Lagrangian does not effect the equations of motion).

Varying $\mathcal{L}_{J}$ gives
\begin{eqnarray}
\label{totAddVariationJ}
k_J\delta\mathcal{L}_{J} & = & \frac{1}{2}\sqrt{|g|}T_{{J}}^{\mu\nu}\delta g_{\mu\nu}+k_J J_\mu \delta j^{\mu}\\
\label{totAddVariationF}
& \overset{\int}{=} & \frac{1}{2}\sqrt{|g|}T_{{J}}^{\mu\nu}\delta g_{\mu\nu}+k_J \nabla_{[\mu} J_{\nu]} \delta f^{\mu\nu}
\end{eqnarray}
or treating $A_\mu$ as independent,
\begin{equation}
\label{totAddVariationA}
\begin{array}{l} 
k_J\delta\mathcal{L}_{J} \overset{\int}{=} \frac{1}{2}\sqrt{|g|}T_{J,A}^{\mu\nu}\delta g_{\mu\nu} + 2 \sqrt{|g|} k_J \nabla_\nu\nabla^{[\mu}J^{\nu]}\delta A_\mu
\end{array}
\end{equation}
\begin{equation}
  \label{jjStressEnergyFromA}
  \begin{array}{ll}
    T_{J,A}^{\mu\nu} \equiv & T_{J}^{\mu\nu} + k_J \left(\nabla_{[\alpha} J_{\beta]} F^{\alpha\beta}g^{\mu\nu} \right.\\
    & \left. - 2 \nabla^{[\mu} J^{\alpha]} F^{\nu\beta} g_{\alpha\beta} - 2 \nabla^{[\nu} J^{\alpha]} F^{\mu\beta} g_{\alpha\beta}\right).
  \end{array}
\end{equation}
The stress-energy tensor of Eq~\ref{jjStressEnergyFromA} can be found in Ref.~\cite{zayats2014} where it was studied in the context of Bopp-Podolsky electrodynamics. It is more complicated than $T_{J}^{\mu\nu}$, but its divergence is fairly simple,
\begin{equation}
\label{divJJTensFromA}
\begin{array}{l} 
  \nabla_\nu T^{\mu\nu}_{J,A} = 2 k_J F^{\mu\nu}\nabla^{\alpha} \nabla_{[\nu} J_{\alpha]},
\end{array}
\end{equation}
which does not satisfy our original guiding condition of $J_\mu \nabla_\nu T_{\rm add}^{\mu\nu}=0$. However, the divergence of the complete stress-energy tensor is
\begin{equation}
\label{eConsFromA}
\begin{array}{l} 
  \nabla_\nu \left(T_{\rm EM}^{\mu\nu}+T_{J,A}^{\mu\nu}\right) = F^{\mu\nu}\left(-J_\nu  + 2 k_J \nabla^\alpha \nabla_{[\nu} J_{\alpha]}\right),
\end{array}
\end{equation}
which is identically perpendicular to the conserved current in parentheses, and hence also has the correct number of independent equations to specify $A^\mu$ (or $J^\mu$).

At this point, it's useful to define a total ``matter'' Lagrangian density (everything except the Lagrangian density for gravity, which we'll add later). The matter Lagrangian with our current terms is
\begin{equation}
\begin{array}{rcl}
\mathcal{L}_{\rm matter} & = & \sqrt{|g|}\left(\frac{1}{4}k_{\rm EM}F_{\mu\nu}F^{\mu\nu}+\frac{1}{2}k_J J_\mu J^\mu \right).
\end{array}
\label{matterLagrangian}
\end{equation}
Note $k_{\rm EM}$ is $+1$ if we're holding $j^\mu$ constant during variation, and $-1$ if we're holding $A_\mu$ constant during variation to obtain the conventional sign of $T_{\rm EM}^{\mu\nu}$ when varying the metric.

The field equations for $A_\mu$ are 
\begin{equation}
\label{fieldEqnsFromA}
\begin{array}{l} 
  \frac{1}{\sqrt{|g|}}\frac{\partial\mathcal{L}_{\rm matter}}{\partial A_\mu} = k_{\rm EM}J^\mu + 2 k_J \nabla_{\alpha} \nabla^{[\mu} J^{\alpha]}=0.
\end{array}
\end{equation}
One thing worth noting is that, for any Lagrangian, conservation of energy-momentum is guaranteed trivially if the field equations resulting from varying all the fields outside of the metric are satisfied, as we'll discuss more in the next section. This is obvious from our current example: if Eq.~\ref{fieldEqnsFromA} is satisfied, then Eq.~\ref{eConsFromA} is trivially satisfied. Thus, from the standpoint of the principle of least action, the field equations fully determine the dynamics, without any help from conservation of energy-momentum.

With this understanding, consider how the field equations differ if we use $j^\mu$ versus $A_\mu$ as independent. The field equations, with $j^\mu$ independent, are 
\begin{equation}
  \frac{\partial\mathcal{L}_{\rm matter}}{\partial j^\mu} = k_{\rm EM}A_\mu + k_J J_\mu = 0,
\label{fieldEqnsFromJ}
\end{equation}
or plugging in the definition for $J^\mu$,
\begin{equation}
 k_{\rm EM}A_\mu + 2 k_J \nabla^\nu \nabla_{[\mu}A_{\nu]} = 0.
 \label{fieldEqnsFromJ2}
\end{equation}
Eq.~\ref{fieldEqnsFromJ2} and Eq.~\ref{fieldEqnsFromA} are of the exact same form, one in terms of $A^\mu$, and the other in terms of $J^\mu$. Both field equations reduce to the Proca equation in flat space-time (as long as one uses the Lorenz gauge for $A^\mu$).

It's also clear, in flat space-time, that any solution to Eq.~\ref{fieldEqnsFromJ2} is a solution to Eq.~\ref{fieldEqnsFromA}. More generally, for any two-form $B_{\mu\nu}$,
\begin{equation}
\begin{array}{l}
  B_{\mu\nu} \delta f^{\mu\nu} \\
  \overset{\int}{=} \sqrt{|g|}\left(\frac{1}{2}B_{\alpha\beta}F^{\alpha\beta}g^{\mu\nu} - 2B^{[\mu\alpha]}F^{\nu\beta}g_{\alpha\beta}\right)\delta g_{\mu\nu}\\
  ~~~~~+2 \sqrt{|g|} \nabla_\nu B^{\mu\nu}\delta A_\mu,
\label{deltaFToDeltaA}
\end{array}
\end{equation}
and for any one form $B_\mu$,
\begin{equation}
\begin{array}{l}
  B_\mu \delta j^\mu \\
  \overset{\int}{=} \sqrt{|g|}\left(\frac{1}{2}\nabla_{[\alpha}B_{\beta]}F^{\alpha\beta}g^{\mu\nu} - 2\nabla^{[\mu}B^{\alpha]}F^{\nu\beta}g_{\alpha\beta}\right)\delta g_{\mu\nu}\\
  ~~~~~+2 \sqrt{|g|} \nabla_\nu\left(\nabla^{[\mu}B^{\nu]}\right)\delta A_\mu.
\label{deltaJToDeltaA}
\end{array}
\end{equation}
So the electromagnetic field equations, treating $A_\mu$ as independent, can be written as
\begin{equation}
  \begin{array}{c}
    2 \nabla^\nu \nabla_{[\mu}B_{\nu]} = 0\\ 
    B_\mu = \frac{\partial\mathcal{L}_{\rm matter}}{\partial j^\mu}.
  \end{array}
\end{equation}
Thus, for any Lagrangian, in flat space-time, any solution to the field equations using $j^\mu$ as independent is also a solution to the field equations using $A_\mu$ as independent. However, the stress-energy tensors differ, and $g_{\mu\nu}$ (gravity) will differ.

Typically space-time curvature is ignored in the study of fundamental particles, due to difficulties in integrating with quantum mechanics, on the grounds that they are light or general relativity may not hold at all for such length scales, or possibly for simplicity. Nevertheless, since the sizes of fundamental particles are smaller than can currently be detected, it's possible space-time curvature can be important to their intradynamics, or even close enough interactions (as one approaches a small enough particle, curvature eventually must be important under the theory of general relativity).

With this in mind, consider what other scalars can be added to the Lagrangian.  Outside of $\mathcal{L}_{\rm EM}$ and $\mathcal{L}_{J}$, there are three other, independent quadratic scalars that depend on $F^{\mu\nu}$ and its first derivatives, two of which are linear in $\eta_{\alpha\beta\mu\nu}$, and thus change sign under a parity transformation (their inclusion would lead to a ``parity violating'' theory)\footnote{There are a few other permutations of indices for first derivative terms, such as $\nabla_{\alpha}F_{\mu\nu}\nabla^{\mu}F^{\alpha\nu}$, but using Maxwell's equations, these reduce to combinations of the terms listed here.}:
\begin{equation}
\begin{array}{rcl}
  \mathcal{L}_{*F} & \equiv & \frac{1}{4}\sqrt{|g|}\eta_{\alpha\beta\mu\nu} F^{\alpha\beta} F^{\mu\nu} \\
  \mathcal{L}_{dF} & \equiv & \frac{1}{4}\sqrt{|g|}\nabla_{\alpha}F_{\mu\nu}\nabla^{\alpha}F^{\mu\nu} \\
  \mathcal{L}_{d*F} & \equiv & \frac{1}{4}\sqrt{|g|}\eta_{\mu\nu\sigma\rho}\nabla_{\alpha}F^{\mu\nu}\nabla^{\alpha}F^{\sigma\rho}.
  \end{array}
\label{otherScalars}
\end{equation}

Using $\df{dd}(\delta \df{A})=0$ (the homogeneous Maxwell's equations, where $\delta$ here is a variation, not the codifferential), the variation of $\mathcal{L}_{*F}$ can be rewritten as\cite{podolsky1942}
\begin{equation}
\begin{array}{rcl}
  \delta \mathcal{L}_{*F} & \overset{\int}{=} & -2 \sqrt{|g|} \nabla_\nu \left(\eta^{\mu\nu\alpha\beta} A_\mu \left(\partial_{[\alpha}\delta A_{\beta]} \right)\right),
  \end{array}
\label{fgNonContrib}
\end{equation}
which is a total divergence, and won't contribute to any variational theory. That's in line with nature, since such a term would certainly be long-range, and there is no long-range parity violation in nature (the weak interaction, which is the only known force to violate parity\cite{wu1957}, is quite short in range).

It does not appear to be possible to write $\mathcal{L}_{dF}$ or $\mathcal{L}_{d*F}$ as a function of $g_{\mu\nu}$ and $j^\mu$, but it is possible to write them as a function of $g_{\mu\nu}$ and $f^{\mu\nu}$, and using Eq.~\ref{deltaFToDeltaA}, convert that to equations of motion using $A_\mu$ as independent. 
First note that, ignoring total divergences, both $\mathcal{L}_{dF}$ and $\mathcal{L}_{d*F}$ can be written as\footnote{Note that $\eta_{\mu\nu\alpha\beta}$ is constant in the sense that its covariant derivative is zero\cite{misner}.}
\begin{equation}
\begin{array}{rcl}
  \mathcal{L}_{dF} & \overset{\int}{=} & -\frac{1}{4}\sqrt{|g|}F^{\mu\nu}\nabla_{\alpha}\nabla^{\alpha}F_{\mu\nu} \\
  \mathcal{L}_{d*F} & \overset{\int}{=} & -\frac{1}{4}\sqrt{|g|}\eta^{\mu\nu\sigma\rho}F_{\mu\nu}\nabla_{\alpha}\nabla^{\alpha}F_{\sigma\rho}.
  \end{array}
\label{firstDerivScalars}
\end{equation}
In flat space-time, $\mathcal{L}_{dF}$ reduces to $\mathcal{L}_J$\cite{gratus2015}, and $\mathcal{L}_{d*F}$ reduces to zero (as we will show shortly). But in the presence of curvature, this is not the case. To show this, it's useful to introduce a Laplacian operator (which differs from $\nabla_\alpha \nabla^\alpha$) that operates on differential forms (antisymmetric tensors), $\df{\Delta}$; This is sometimes called the ``Hodge Laplacian'', ``de Rham wave operator'', or ``Laplace-de Rham operator''\cite{deRham1955,misner}, 
\begin{equation}
\begin{array}{rcl}
  \df{\Delta} & \equiv & (\df{d} + \df{\delta})^2 = (\df{d\delta} + \df{\delta d}) \\
  & = & \df{d*d*} + \df{*d*d}. \\
  \end{array}
\label{deltaDef}
\end{equation}
As it turns out, in flat space-time $\df{\Delta}=-\nabla_\alpha \nabla^\alpha$, but not so in curved space-time. In curved space-time, extra terms appear proportional to the curvature (these extra terms were first derived by Weitzenbock\cite{deRham1955}).
 Operating on any two-form, $\df{B}$, the relation is\cite{deRham1955}
\begin{equation}
\begin{array}{rcl}
  (\df{\Delta B})_{\mu\nu} & + & \nabla_\alpha \nabla^\alpha B_{\mu\nu} = W^{\alpha\beta}_{~~~\mu\nu}B_{\alpha\beta} \\
  W_{\alpha\beta\mu\nu} & \equiv & R_{\alpha[\mu}g_{\nu]\beta} - R_{\beta[\mu}g_{\nu]\alpha} - R_{\alpha\beta\mu\nu}.
  \end{array}
\label{weitzenbock2form}
\end{equation}
Also, the current 1-form, $\df{J} \equiv \df{*d*F} = \df{\delta F}$, so 
\begin{equation}
\begin{array}{rcl}
  \df{\Delta F} & = & \df{dJ}.
  \end{array}
\label{laplacianF}
\end{equation}
Using these relationships, we find
\begin{equation}
\begin{array}{rcl}
  \mathcal{L}_{dF} & \overset{\int}{=} & \frac{1}{4}\sqrt{|g|}F^{\mu\nu}\left[(\df{\Delta F})_{\mu\nu}-W^{\alpha\beta}_{~~~\mu\nu}F_{\alpha\beta}\right] \\
  & = & \frac{1}{4}\sqrt{|g|}F^{\mu\nu}\left[(\df{dJ})_{\mu\nu}-W^{\alpha\beta}_{~~~\mu\nu}F_{\alpha\beta}\right] \\
  \mathcal{L}_{d*F} & \overset{\int}{=} & \frac{1}{4}\sqrt{|g|}\eta^{\mu\nu\sigma\rho}F_{\mu\nu}\left[(\df{\Delta F})_{\sigma\rho}-W^{\alpha\beta}_{~~~\sigma\rho}F_{\alpha\beta}\right] \\
  & = & \frac{1}{4}\sqrt{|g|}\eta^{\mu\nu\sigma\rho}F_{\mu\nu}\left[(\df{dJ})_{\sigma\rho}-W^{\alpha\beta}_{~~~\sigma\rho}F_{\alpha\beta}\right].
  \end{array}
\label{firstDerivScalars2}
\end{equation}
One also finds
\begin{equation}
  \begin{array}{rcl}
    \sqrt{|g|}F^{\mu\nu}(\df{dJ})_{\mu\nu} & \overset{\int}{=} & 2 \sqrt{|g|} J_\mu J^\mu \\
    \sqrt{|g|}\eta^{\mu\nu\sigma\rho}F_{\mu\nu}(\df{dJ})_{\sigma\rho} & \overset{\int}{=} & 0,
  \end{array}
  \label{firstDerivScalars3}
\end{equation}
the latter being zero as it reduces to terms involving $\df{dF}=\df{ddA}=0$ (the homogeneous Maxwell's equations) and total divergences. Thus, we can rewrite $\mathcal{L}_{dF}$ and $\mathcal{L}_{d*F}$ as\footnote{We write $X_{\alpha\beta\mu\nu}$ in this fashion to guarantee it is symmetric under interchange of the first pair and second pair of indices, since it is only that portion that contributes here. Note that $W_{\alpha\beta\mu\nu}$ has this property inherently.}
\begin{equation}
\begin{array}{rcl}
  \mathcal{L}_{dF} & \overset{\int}{=} & \sqrt{|g|}\left(\frac{1}{2} J_\mu J^\mu - \frac{1}{4}W_{\alpha\beta\mu\nu}F^{\alpha\beta}F^{\mu\nu}\right) \\
  \mathcal{L}_{d*F} & \overset{\int}{=} & - \frac{1}{4}\sqrt{|g|}X_{\alpha\beta\mu\nu}F^{\mu\nu}F^{\alpha\beta}. \\
  X_{\alpha\beta\mu\nu} & \equiv & \frac{1}{2}\left(\eta_{\alpha\beta\sigma\rho}W_{\mu\nu}^{~~~\sigma\rho} + \eta_{\mu\nu\sigma\rho}W_{\alpha\beta}^{~~~\sigma\rho}\right).
  \end{array}
\label{firstDerivScalars4}
\end{equation}
The Lagrangian densities of Eq.~\ref{firstDerivScalars4} are examples of ``non-minimally coupled'' Lagrangians studied in Ref.~\cite{balakin2005} and subsequent publications in recent years. In those publications, they study different products of curvature and $F^{\mu\nu}$, choosing which terms to include and their relative strengths in an {\em ad hoc} or phenomenological manner\cite{balakin2005}; here we have arrived at relatively few options by including all quadratic products of first-order derivatives of $F^{\mu\nu}$. The variation of these Lagrangian densities may be written as\footnote{The variations of $W_{\alpha\beta\mu\nu}$ and $X_{\alpha\beta\mu\nu}$ with respect to the metric are more complicated, and don't contribute much to the current discussion, so these will be left for a later publication.},
\begin{equation}
\begin{array}{rcl}
  \delta \mathcal{L}_{dF} & \overset{\int}{=} & \delta \left(\frac{1}{2} \sqrt{|g|} J_\mu J^\mu \right) \\
  & & - \frac{1}{4}\left[\frac{\partial \left(\frac{1}{\sqrt{|g|}} W_{\alpha\beta\mu\nu}\right)}{\partial g_{\sigma\rho}} f^{\alpha\beta}f^{\mu\nu}\right]\delta g_{\sigma\rho} \\
  & & - \frac{1}{2} W_{\alpha\beta\mu\nu}F^{\alpha\beta}\delta f^{\mu\nu} \\
  \delta \mathcal{L}_{d*F} & \overset{\int}{=} & - \frac{1}{4}\left[\frac{\partial \left( \frac{1}{\sqrt{|g|}}X_{\alpha\beta\mu\nu}\right)}{\partial g_{\sigma\rho}} f^{\alpha\beta}f^{\mu\nu}\right]\delta g_{\sigma\rho} \\
  && - \frac{1}{2} X_{\alpha\beta\mu\nu}F^{\alpha\beta}\delta f^{\mu\nu}.
  \end{array}
\label{firstDerivsPartialA}
\end{equation}
From this and Eq.~\ref{deltaFToDeltaA},
\begin{equation}
\begin{array}{rcl}
  \frac{\partial \mathcal{L}_{dF}}{\partial A_\mu} & = & \sqrt{|g|} \left[ 2 \nabla_\nu \nabla^{[\mu}J^{\nu]} - \nabla_\nu \left(W^{\alpha\beta\mu\nu}F_{\alpha\beta}\right)\right] \\
  \frac{\partial \mathcal{L}_{d*F}}{\partial A_\mu} & = & - \sqrt{|g|} \nabla_\nu\left(X^{\alpha\beta\mu\nu}F_{\alpha\beta}\right).
  \end{array}
\label{firstDerivVariation}
\end{equation}

We now have all the pieces to treat the most general matter Lagrangian that includes only the degrees of freedom in $F^{\mu\nu}$, including all quadratic terms involving $F^{\mu\nu}$ and its first derivatives:\footnote{To emphasize we've made no assumption on $J^\mu$, we've replaced it with its definition here, $J^\mu \equiv \nabla_\nu F^{\mu\nu}$.}
\begin{equation}
\begin{array}{l}
  \mathcal{L}_{\rm matter} = \\
  ~~- k_{\rm EM} \mathcal{L}_{\rm EM} + k_J \mathcal{L}_{J} + k_{dF}\left(\mathcal{L}_{dF} - \mathcal{L}_{J}\right) + k_{d*F} \mathcal{L}_{d*F} \\
   =  \frac{1}{4} k_{\rm EM} \sqrt{|g|}F_{\mu\nu}F^{\mu\nu} \\
   ~~ + \frac{1}{2} k_J \sqrt{|g|}\nabla_\nu F^{\mu\nu} \nabla^\alpha F_{\mu\alpha} \\
   ~~ + \frac{1}{4} k_{dF} \sqrt{|g|}\left(\nabla_\alpha F^{\mu\nu} \nabla^\alpha F_{\mu\nu} - 2 \nabla_\nu F^{\mu\nu} \nabla^\alpha F_{\mu\alpha}\right) \\
   ~~ + \frac{1}{4} k_{d*F} \sqrt{|g|}\eta_{\mu\nu\sigma\rho}\nabla_\alpha F^{\mu\nu} \nabla^\alpha F^{\sigma\rho},
  \end{array}
\label{completeMatterL}
\end{equation}
where we've subtracted $\mathcal{L}_J$ from $\mathcal{L}_{dF}$ to isolate the curvature dependent portion of $\mathcal{L}_{dF}$; the contribution of $\mathcal{L}_{dF} - \mathcal{L}_{J}$ to the total Lagrangian is proportional to curvature by Eq.~\ref{firstDerivScalars4}. It's not apparent how to write $\mathcal{L}_{\rm matter}$ in terms of $j^\mu$, which precludes varying it with respect to $j^\mu$, but varying $A_\mu$, the electromagnetic field equations are
\begin{equation}
\begin{array}{l}
  \frac{1}{\sqrt{|g|}}\frac{\partial\mathcal{L}_{\rm matter}}{\partial A_\mu} = \\ k_{\rm EM}J^\mu
  + k_J J_{dJ}^\mu
  + k_{dF} J_{dF}^\mu
  + k_{d*F} J_{d*F}^\mu
  =0,
  \end{array}
\label{fieldEquationsOfJs}
\end{equation}
where each term is comprised of a conserved current,
\begin{equation}
  \begin{array}{rcl}
    J_{dJ}^\mu & \equiv & 2 \nabla_{\alpha} \nabla^{[\mu} J^{\alpha]} \\
    J_{dF}^\mu & \equiv & - \nabla_\nu \left(W^{\alpha\beta\mu\nu}F_{\alpha\beta}\right) \\
    J_{d*F}^\mu & \equiv & - \nabla_\nu\left(X^{\alpha\beta\mu\nu}F_{\alpha\beta}\right),
  \end{array}
\label{newConservedCurrents}
\end{equation}
with $\nabla_\mu J_{dJ}^{\mu}=\nabla_\mu J_{dF}^{\mu}=\nabla_\mu J_{d*F}^{\mu}=0$ being identically true by virtue of being a divergence of an anti-symmetric 2-tensor\cite{misner}. In fact, varying any Lagrangian term, that can be written in terms of $f^{\mu\nu}$ and $g_{\mu\nu}$, with respect to $A_\mu$, will result in such a conserved current by Eq.~\ref{deltaFToDeltaA}. This means given initial conditions on a space-like surface, one of the four field equations will be determined by the other three; this will preserve the gauge freedom in $A_\mu$ in solutions.

To include gravity, add the Einstein-Hilbert Lagrangian, $\mathcal{L}_{\rm EH}$, to give the total Lagrangian\cite{misner}
\begin{equation}
\begin{array}{rcl}
  \mathcal{L}_{\rm total} & = & \mathcal{L}_{\rm EH} + \mathcal{L}_{\rm matter}\\
  \mathcal{L}_{\rm EH} & \equiv & \frac{1}{16\pi}\sqrt{|g|}R,
\end{array}
\label{totalLagrangian}
\end{equation}
where $R$ is the scalar curvature. The variation of this Lagrangian is\footnote{A derivation of the variation of the scalar curvature can be found in \cite{dirac1975}.}
\begin{equation}
\begin{array}{rcl}
\delta \mathcal{L}_{\rm total} & = & \sqrt{|g|}\left[-\frac{1}{16\pi}(R^{\mu\nu}-\frac{1}{2}R g^{\mu\nu}) + \frac{1}{2}T^{\mu\nu}\right]\delta g_{\mu\nu} \\
& & +\frac{\partial\mathcal{L}_{\rm matter}}{\partial A_\mu}\delta A_\mu \\
T^{\mu\nu} & = & \frac{2}{\sqrt{|g|}}\frac{\partial\mathcal{L}_{\rm matter}}{\partial g_{\mu\nu}},
\end{array}
\label{totalVariation}
\end{equation}
where $R^{\mu\nu}$ is the Ricci curvature tensor (or $R^{\mu\nu}-\frac{1}{2}R g^{\mu\nu}$ is the divergence-free Einstein tensor).
Setting what multiplies $\delta g_{\mu\nu}$ to zero gives Einstein's equations, with our new stress-energy tensor\footnote{Performing the variation of the new terms with respect to $g_{\mu\nu}$ to calculate the form of $T^{\mu\nu}$ is left for a future publication, as it is somewhat complicated. In the next section, however, we'll calculate $\nabla_\mu T^{\mu\nu}$}:
\begin{equation}
\begin{array}{rcl}
R^{\mu\nu}-\frac{1}{2}R g^{\mu\nu} = 8\pi T^{\mu\nu}.
\end{array}
\label{completeEinstein}
\end{equation}

To summarize, removing the point-charge assumption, and including all possible quadratic Lagrangians in terms of $F^{\mu\nu}$ and its first derivatives, we find a self-consistent, gauge invariant theory compatible with general relativity\footnote{Self-consistent in the sense that energy-momentum can actually be conserved in a self-consistent way, without requiring $J^\mu=0$ (which is what is required with just $\mathcal{L}_{\rm EM}$).}. This theory, introduces an internal force (from the current-current interaction, $J_\mu J^\mu$), and two new very short range forces mediated by space-time curvature, one which will result in a parity violating theory.

We've taken a somewhat round-about method of getting to the theory of Eq.~\ref{totalLagrangian} and Eq.~\ref{completeMatterL}. We started with trying to find ways to self-consistently finding stress-energy tensors, requiring there only be 3 independent equations of motion, which led us to the matter Lagrangian of Eq.~\ref{matterLagrangian}, using $j^\mu$ as independent rather than $A_\mu$. While this theory is self-consistent, treating $j^\mu$ as independent does not appear to allow for the new curvature-mediated, short-range forces (and hence cannot violate parity).

An interesting fact that came from this is we obtained enough equations of motion directly from conservation of energy-momentum, and we could solve those to obtain solutions, rather than using the field equations from the principle of least action. In the next section, we find that by modifying the principle of least action, we can develop variational theories where energy-momentum conservation are the fundamental field equations.

\subsection{\label{sec:diffeo}Least Action, Energy-Momentum Conservation, and Diffeomorphism Invariance}

Conventionally, conservation of energy-momentum equations are found by using principle of least action, Eq.~\ref{deltaActionEqualsZero}, including gravity (the Einstein-Hilbert Lagrangian), which results in Einstein's equations with the stress-energy tensor, $T^{\mu\nu}$, the divergence of which must be zero, since it equals the Einstein tensor, which has zero divergence by the Bianchi identities\cite{misner}\footnote{Or alternatively, one finds the canonical stress-energy tensor, and then manipulates it to agree with the symmetric $T^{\mu\nu}$ found from varying $g_{\mu\nu}$.\cite{jackson}.}.

Using a variational method, is there a way to make energy-momentum conservation fundamental, where those are the equations that fully determine the fields? To do this, rather than allow independent fields to vary arbitrarily, we will restrict their variations somewhat.

For each independent tensor in the Lagrangian, ${{T_i}^{\mu_1, \mu_2...}}_{\nu_1,\nu_2...}$, let $\xi_i^\mu$ be a smooth vector field with compact support, and let $\phi_\lambda$ be the one parameter family of diffeomorphisms, which is the global flow (or the one parameter group action) associated with $\xi_i^\mu$, where $\lambda$ is a real number\cite{lee2012}. The change of a tensor induced by this flow, which we will denote as $\overline{\delta}_\lambda$, is 
\begin{equation}
\begin{array}{rcl}
  \multicolumn{3}{l}{
    \overline{\delta}_{\lambda}{{T_i}^{\mu_1, \mu_2...}}_{\nu_1,\nu_2...}
  }\\
  & \equiv & \phi_{\lambda *}\left[{{T_i}^{\mu_1, \mu_2...}}_{\nu_1,\nu_2...}(x'^{\mu})\right] - {{T_i}^{\mu_1, \mu_2...}}_{\nu_1,\nu_2...}(x^{\mu})\\
x'^{\mu} & = & \phi_\lambda(x^\mu) \\
\end{array}
\label{infDiffeoCoord}
\end{equation}
where $\phi_{\lambda *}$ is the pullback of $T_i$ by $\phi_\lambda$, using the convention of Carroll\cite{carroll1997}. In the limit of $\lambda\rightarrow 0$, to first order in $\lambda$,\cite{carroll1997},
\begin{equation}
\begin{array}{rcl}
  \overline{\delta}_\lambda{{T_i}^{\mu_1, \mu_2...}}_{\nu_1,\nu_2...} & \approx & \overline{\delta}{{T_i}^{\mu_1, \mu_2...}}_{\nu_1,\nu_2...} \\
  \overline{\delta}{{T_i}^{\mu_1, \mu_2...}}_{\nu_1,\nu_2...} & \equiv & \lambda \pounds_{\xi_i}{{T_i}^{\mu_1, \mu_2...}}_{\nu_1,\nu_2...}
\end{array}
\label{diffeoLie}
\end{equation}
where $\pounds_{\xi_i}$ is the Lie derivative. If we replace $\delta{{T_i}^{\mu_1, \mu_2...}}_{\nu_1,\nu_2...}$ with $\overline{\delta}{{T_i}^{\mu_1, \mu_2...}}_{\nu_1,\nu_2...}$ in the principle of least action, Eq.~\ref{deltaActionEqualsZero}, becomes
\begin{equation}
  \begin{array}{rcl}
    \overline{\delta} S & = & \int \frac{\partial \mathcal{L}}{\partial {{T_i}^{\mu_1, \mu_2...}}_{\nu_1,\nu_2...}}\overline{\delta}{{T_i}^{\mu_1, \mu_2...}}_{\nu_1,\nu_2...} d^4 x = 0 \\
    & = & \int \frac{\partial \mathcal{L}}{\partial {{T_i}^{\mu_1, \mu_2...}}_{\nu_1,\nu_2...}} \lambda \pounds_{\xi_i}{{T_i}^{\mu_1, \mu_2...}}_{\nu_1,\nu_2...} d^4 x = 0,
  \end{array}
  \label{diffeoActionEqualsZero}
\end{equation}
where each $\xi_i^\mu$ is associated with an independent field, ${{T_i}^{\mu_1, \mu_2...}}_{\nu_1,\nu_2...}$. Each $\xi_i^\mu$ is independent from the others, and each is allowed to vary arbitrarily on the manifold. This may be viewed as allowing each independent field to vary due to an arbitrary, infinitesimal active coordinate transformation, with each independent field being allowed to transform independently from the others. This produces the following equations of motion for each tensor field:
\begin{equation}
  \begin{array}{rcl}
    \frac{\partial \mathcal{L}}{\partial {{T_i}^{\mu_1, \mu_2...}}_{\nu_1,\nu_2...}} \pounds_{\xi_i}{{T_i}^{\mu_1, \mu_2...}}_{\nu_1,\nu_2...} \overset{\int}{=} 0.
  \end{array}
  \label{diffeoFieldEquations}
\end{equation}
$\pounds_{\xi_i}{{T_i}^{\mu_1, \mu_2...}}_{\nu_1,\nu_2...}$ will have terms that involve $\xi_i^\mu$ and its first derivatives. Any derivative of $\xi_i^\mu$ can be removed by integration by parts (where the compact support of $\xi_i^\mu$ will remove any boundary terms). Then, since $\xi_i^\mu$ is arbitrary, whatever multiplies it must be zero. This provides 4 equations of motion per point in space-time.

The variation of the metric due to the diffeomorphism of Eq.~\ref{infDiffeoCoord} is\cite{carroll1997}
\begin{equation}
  \overline{\delta}g_{\mu\nu} = \lambda \pounds_{\xi_g} g_{\mu\nu} = 2 \lambda \nabla_{(\mu}\xi_{g,\nu)},
  \label{metricVariation}
\end{equation}
where $\xi_g^\mu$ is the $\xi_i^\mu$ associated with $g_{\mu\nu}$. This makes the field equations for the metric
\begin{equation}
\begin{array}{rcl}
\frac{\partial\mathcal{L}}{\partial g_{\mu\nu}}\overline{\delta} g_{\mu\nu} & = & 2 \frac{\partial\mathcal{L}}{\partial g_{\mu\nu}}\lambda \nabla_{(\mu}\xi_{{\rm g},\nu)} \overset{\int}{=} 0.
\end{array}
\label{dMetricDiffeoEqns}
\end{equation}
Integrating by parts to move the derivative from $\xi_g^\mu$ to $\frac{\partial \mathcal{L}}{\partial g_{\mu\nu}}$ gives conservation of total energy-momentum directly from the principle of diffeomorphism invariance\cite{carroll1997}.

What of other fields? The Lie derivative of $A_\mu$ is\cite{carroll1997}
\begin{equation}
\begin{array}{l}
  \pounds_{\xi_A} A_\mu = \xi_A^\nu \nabla_\nu A_\mu + A_\nu \nabla_\mu \xi_A^\nu,
\end{array}
\label{deltaFEq}
\end{equation}
and, for any vector $B^\mu$, one finds
\begin{equation}
\begin{array}{l}
  \sqrt{|g|}B^\mu\overline{\delta} A_\mu \overset{\int}{=} \sqrt{|g|}\left(-A_\mu \nabla_\nu B^\nu +B^\nu F_{\mu\nu}\right)\lambda \xi_A^\mu,
\end{array}
\label{diffeoAEq}
\end{equation}
where $\overset{\int}{=}$ is defined by Eq.~\ref{intEquals}. We've already established that any $\frac{\partial\mathcal{L}}{\partial A_\mu}$ will be a conserved current density if $\mathcal{L}$ can be written as a function of $F^{\mu\nu}$ and its derivatives. If this is the case, and $\sqrt{|g|}B^\mu = \frac{\partial\mathcal{L}}{\partial A_\mu}$, the first term on the right hand side in Eq.~\ref{diffeoAEq} will be zero. 

As a simple example, consider if we use Eq.~\ref{matterLagrangian} as our matter Lagrangian, which using the principle of least action results in the fields equations (varying $A_\mu$) of Eq.~\ref{fieldEqnsFromA}. Using those field equations as $B^\mu$ in Eq.~\ref{diffeoAEq}, one obtains
\begin{equation}
\label{diffeoEqnsFromA}
\begin{array}{l} 
  \frac{\partial \mathcal{L}}{\partial A_\mu} \overline{\delta} A_\mu \overset{\int}{=} \sqrt{|g|}\left(k_{\rm EM}J^\nu + 2 k_J \nabla_{\alpha} \nabla^{[\nu} J^{\alpha]}\right)F_{\mu\nu}\lambda \xi_A^\mu \overset{\int}{=} 0.
\end{array}
\end{equation}
Requiring this to be zero for arbitrary $\xi_A^\mu$ leads exactly to the conservation of energy-momentum from the stress-energy tensor derived from $\frac{\partial \mathcal{L}}{\partial g_{\mu\nu}}$! We arrived at energy-momentum conservation without varying the metric to find $T^{\mu\nu}$ and taking its divergence; we never varied the metric at all. 

This redundancy between energy-momentum conservation arising from the $\overline{\delta} g_{\mu\nu}$ variation, and the $\overline{\delta} A_\mu$ variation, is no coincidence. The variation of {\em any} scalar density (such as a Lagrangian density) due to an arbitrary infinitesimal diffeomorphism is a total divergence, zero for our purposes:
\begin{equation}
\begin{array}{rcl}
\overline{\delta}(\sqrt{|g|}L) & = & \sqrt{|g|}\left(\frac{1}{2} L g^{\mu\nu}\overline{\delta} g_{\mu\nu}+\overline{\delta} L\right) \\ 
 & = & \sqrt{|g|}\left(L g^{\mu\nu} \lambda \nabla_{\mu}\xi_{\nu}+\lambda \xi^\mu \nabla_\mu L\right) \\ 
 & = & \lambda\sqrt{|g|}\nabla_\mu\left(L \xi^\mu \right).
\end{array}
\label{diffeoScalarZero}
\end{equation}
If a Lagrangian is comprised of two independent fields, then their $\overline{\delta}$ variations will {\em always} be exactly the same with opposite sign. In the case of the metric and one other field, in this example $A_\mu$, the field equation from the variation due to the other field is guaranteed to be the same as energy-momentum conservation obtained from $\nabla_\mu T^{\mu\nu}=0$. If more fields are added to the theory, for each non-metric field in the theory, a separate conservation of energy-momentum equation is generated for that field, and by Eq.~\ref{diffeoScalarZero}, the sum of all the individual conservation of energy-momentum equations add to $\nabla_\mu T^{\mu\nu}=0$\footnote{Eq.~\ref{diffeoScalarZero} provides a useful consistency check for calculating variations. If the sum of all of the variations are not zero, some mistake was made.}. Note this means a theory generated using the diffeomorphism approach of Eq.~\ref{diffeoActionEqualsZero} with a single field (i.e. gravity with no matter, or electromagnetism with no gravity) is trivially zero. It takes at least two independent fields (or in the context of general relativity, gravity plus at least one matter field) to generate a nontrivial theory.

Using this methodology, sometimes we can find the conservation of energy-momentum equations much more easily than finding $T^{\mu\nu}$ and taking its divergence. A good example is the theory including all first-derivative terms of $F^{\mu\nu}$, Eq.~\ref{completeMatterL}. From Eq.~\ref{diffeoAEq}, energy-momentum conservation is always just $F^{\mu\nu}$ contracted with the field equations from the principle of least action (if $\frac{\partial \mathcal{L}}{\partial A_\mu}$ is a conserved current):
\begin{equation}
\begin{array}{c}
  F_{\mu\nu}\left[k_{\rm EM}J^\mu 
   +  k_J J_{dJ}^\mu + k_{dF} J_{dF}^\mu
  + k_{d*F} J_{d*F}^\mu\right]
  =0,
  \end{array}
\label{eConsEquationsOfJs}
\end{equation}
and energy-momentum conservation for this theory amounts to the electromagnetic field delivering energy to all the conserved currents in Eq.~\ref{fieldEquationsOfJs}; this energy-momentum into and out of these currents must balance. This gives only 3 equations of motion by virtue of the equations of motion being perpendicular to the total current in brackets.

What one does not obtain from this variational principle is Einstein's equations. To obtain a theory that fully specifies the metric as well as any additional vector fields, we could use the above methodology for non-metric fields, but to obtain Einstein's equations one would need to vary $g_{\mu\nu}$ arbitrarily as in the conventional principle of least action. Or one could simply add Einstein's equations to the theory. Both options seem less than satisfactory; perhaps there is another way to argue that Einstein's equations arise directly from the principle of diffeomorphism invariance. In any case, if we add Einstein's equations, as long as all the non-metric fields are vector fields (with at most 4 degrees of freedom), one can use this method to find solutions, where the fundamental equations of motion for the non-metric fields are energy-momentum conservation equations. 

It's worth noting that for a given Lagrangian, any solution to the least action principle field equations (the typical Euler-Lagrange field equations) will also be a solution to the conservation of energy-momentum field equations found from diffeomorphism invariance; this is obvious from the fact that the least action equations are true for arbitrary field variations, and all we've done is restricted the variation. However, there may be solutions to Eq.~\ref{diffeoActionEqualsZero}, which aren't solutions to the associated least action theory (we'll show one in the next section).


\section{Solutions to Equations of Motion}

To the author's knowledge, for the first time, we have classical electromagnetic theories from which we can, in a self-consistent way, attempt to find stable, non-point-charge solutions. If stable solutions exist, then we can calculate the mass and charge of such a fundamental object. In this section, we turn our attention to such possible stable solutions.  

\subsection{Superluminal (Space-Like) Currents}

We'll restrict ourselves to static, spherically symmetric solutions in this paper, but we would be remiss not to address that the fundamental particles of nature appear to have the property that they spin faster than the speed of light. If they are, in fact, extended objects, this would amount to space-like electromagnetic current being present in solutions. If a theory is to be able to describe such particles, it must have this ability.

The energy of a particle (or any compact object) certainly diverges as its center-of-mass velocity approaches the speed of light (see Eq.~\ref{particleEqOfMotion2} for example). Most fluids also share this property: the stress-energy tensor diverges as $u^\mu$ approaches being light-like\cite{bonazzola1993}. This divergence prevents fluids' bulk velocity from achieving or exceeding the speed of light (becoming space-like). However, with a little inspection, any theory that is written purely in terms of positive powers of $F^{\mu\nu}$ and its derivatives does not preclude light-like or space-like currents, since none of the quantities in $\mathcal{L}$ or $T^{\mu\nu}$ diverge as $J^\mu$ goes from time-like to light-like to space-like.

Although the current may be well behaved as its bulk velocity approaches $c$, one may ask whether the energy in its field diverges. Point charges (or any discrete body of charge), for instance, have a strict speed limit of the speed of light, because the electromagnetic field (and the energy and momentum of the field) diverges as the speed approaches $c$\cite{jackson}. However, this is not true for currents in general. There is nothing in the electromagnetic fields of a continuous current $J^\mu$ preventing it from being space-like, or changing from time-like to light-like to space-like. This can easily be seen by taking the fields of an infinite wire with continuous charge density $\rho$, which is constant in time. Now say $\vec{J}$ increases linearly with time from 0. The electric and magnetic field simply change linearly in time, while at some point $|\vec{J}|$ equals $\rho$, and at later times is greater than $\rho$.





Note this in no way violates causality. The equations are fully covariant. For our perfect fluid example, the speed at which perturbations in the fluid travel, the speed of sound in the fluid, is in fact the speed of light ($v_s^2=dP/d\epsilon=1$\cite{tooper1964}; $c=1$ in our units). Of course, if bound charge distributions exist, they will have the familiar center-of-mass equations of motion, Eq.~\ref{particleEqOfMotion2}, where the velocity of the resulting charged object is limited by the speed of light. 

\subsection{Solutions in Flat Space-Time}

In flat space-time, the Lagrangian of Eq.~\ref{completeMatterL} reduces to Eq.~\ref{matterLagrangian}. We've already mentioned that in flat space-time, the equations of motion associated with the Lagrangian, Eq.~\ref{matterLagrangian}, are the Proca equation (if we treat $j^\mu$ as independent, this is in terms of $A^\mu$; if we treat $A^\mu$ as independent, it is in terms of $J^\mu$). The solutions of the Proca equation are well studied, and we won't delve into them more deeply here, except to say: it's impossible in flat space-time to find a stable solution to these equations that will quantize mass and charge\footnote{One might find classical solutions and then quantize the solutions after the fact, as in string theory, but this is beyond the scope of this paper.}.  This is because the field equations are linear in $A^\mu$ and its derivatives, and thus any multiple of a solution will be a solution. Even from the standpoint of energy-momentum conservation, the equations are purely quadratic in $A^\mu$ and its derivatives, and any multiple of a solution will still be a solution. Some mixed non-linearity would be necessary to produce any hope of quantization. General relativity is a very nonlinear theory, and as we'll see, curved space-time provides exactly that kind of nonlinearity necessary to produce quantized solutions.

\subsection{Curved Space-Time and Mass/Charge Quantization}

Due to the nonlinearity of Einstein's equations, finding solutions is no simple task. Because of this, to start, we will use the simplest theory, the Lagrangian of Eq.~\ref{matterLagrangian}, treating $j^\mu$ as independent, yielding $T_J^{\mu\nu}$ as our addition to the stress-energy tensor. Its relationship to a perfect fluid yields methods for finding solutions that are readily available. Then, we will address the Lagrangian of Eq.~\ref{matterLagrangian} treating $A_\mu$ as independent. Finally, we'll study the strengths of the terms in the Lagrangian of Eq.~\ref{completeMatterL} treating $A_\mu$ as independent. Solutions to Eq.~\ref{completeMatterL} would be more interesting, since it has parity violation for instance, but due to their complexity, are left for future study (partially because it appears no static, spherical solutions exist). 

\subsubsection{\label{sec:solutionJConst}Lagrangian of Eq.~\ref{matterLagrangian} with $j^\mu$ as Independent}

We'll restrict ourselves to spherically symmetric, time-independent situations, using coordinates, $(t,r,\theta,\phi)$. In this case, $\vec{J}$ is zero. Also, near the center of any spherical charge distribution with finite charge, the radial component of the electric field limits to one power of $r$ greater than the lowest power of $r$ in $\rho$ (see Gauss' law). Therefore, for a small enough distribution (or close enough to its center), the contribution to the stress-energy tensor from the $T_{\rm EM}^{\mu\nu}$ is negligible compared to the contribution from $T_J^{\mu\nu}$. Thus near the origin, for a spherically symmetric static charged object, the stress-energy tensor may be approximated as that of a perfect fluid with the relationship between the energy density, $\epsilon$, and pressure, $P$, as $\epsilon=P\approx\frac{k_J}{2}\rho^2$ (its equation of state).

The ``Tolman V'' solution with $n=1$ and $R\rightarrow\infty$\cite{tolman1939} is the spherically symmetric solution for the case of $\epsilon=P$. The energy and pressure of that solution is
\begin{equation}
\epsilon=P=\frac{1}{16 \pi}\frac{1}{r^2}, 
\label{tolmanSoln}
\end{equation}
which makes the charge density
\begin{equation}
\rho=\frac{1}{\sqrt{8 \pi k_J}}\frac{1}{r}. 
\label{tolmanRhoSoln}
\end{equation}
The metric for this distribution (again ignoring the electric field) corresponds to the line element\cite{tolman1939}
\begin{equation}
ds^2=-\left(\frac{r}{r_1}\right)^2 dt^2+2dr^2+r^2d\theta^2+r^2\sin^2\theta d\phi^2,
\label{tolmanMetric}
\end{equation}
where $r_1$ is an arbitrary constant. The charge and energy densities near the origin are singular, but with a finite volume integral out to a finite $r$. However, the integrals of the energy and charge out to $r=\infty$ are not finite, and the metric never approaches an asymptotically flat form. This metric also cannot smoothly connect to an external metric (such as the Schwarzschild metric\cite{schwarzschild1916}) since the pressure never reaches zero.

It's interesting to note that using $A_\mu$ as independent with $\mathcal{L}_{\rm matter}= \mathcal{L}_J$ produces the same solution: plugging this solution into $T_{J,A}^{\mu\nu}$, one finds $T_{J,A}^{\mu\nu}=T_J^{\mu\nu}$, so in this case, there is no difference between treating $j^\mu$ or $A_\mu$ as independent.

\begin{figure}[h]
{
\includegraphics[scale=0.65]{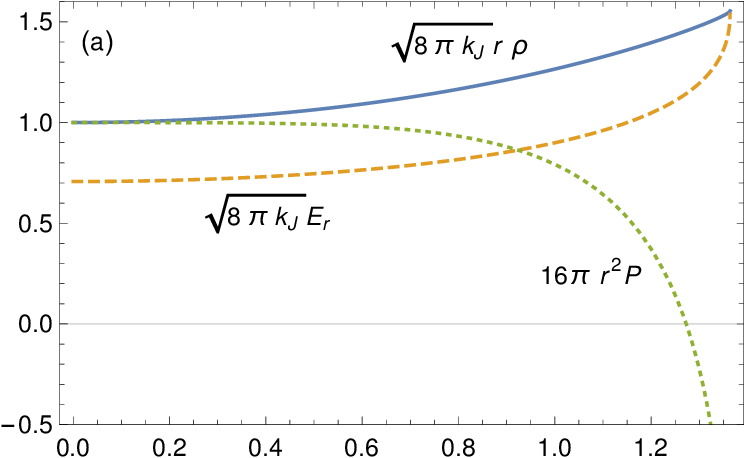}\\
\hspace*{0.4cm}\includegraphics[scale=0.62]{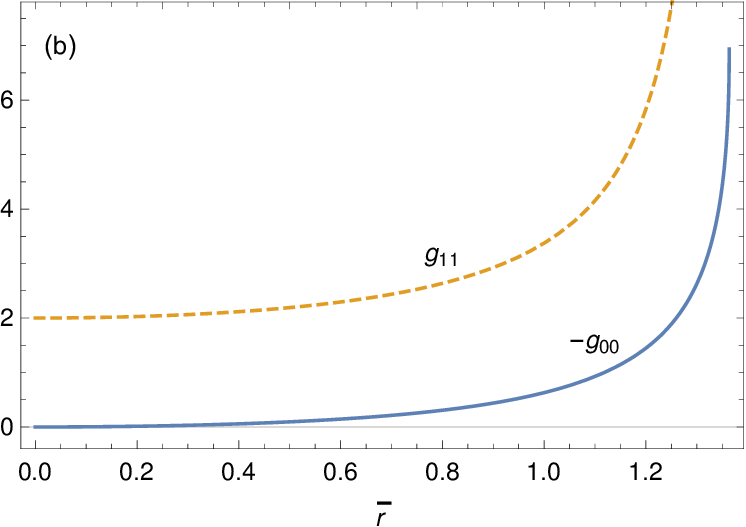}
\caption{\label{fig:chargeFieldMetric} (a) The charge density ($\rho=\sqrt{|g_{00}|}J^0$, solid blue curve), radial electric field ($E_r=\sqrt{|g_{00}|g_{11}}F^{01}$, dashed orange curve), and pressure ($P=T_1^1$, dotted green curve) as a function of $\overline{r}$; (b) the time ($-g_{00}$, solid blue curve) and radial ($g_{11}$, dashed orange curve) metric components. Note $\overline{r} = r/\sqrt{k_J}$ to make it unitless; The variables have been scaled such that all curves are independent of $k_J$, and $\rho$ and $P$ are shown scaled to the value they would have with no electric field. $g_{00}$ is arbitrary to a multiplicative constant, but all other curves do not depend on this constant.}
}
\end{figure}

We can now use the solution Eq.~\ref{tolmanRhoSoln} and Eq.~\ref{tolmanMetric} to describe the small-$r$ behavior (as a boundary condition at small $r$) where the electric field is negligible. To do this, we assume Eq.~\ref{tolmanRhoSoln} and Eq.~\ref{tolmanMetric} are the solutions for $0 \leq r \leq 10^{-6} \sqrt{k_J}$. Then we numerically integrate Einstein's equations (using Mathematica) {\em including} the electric field from $r = 10^{-6} \sqrt{k_J}$ until the numerical solver can no longer continue\footnote{While Tolman et al. assume a perfect fluid, they introduce equations in terms of the more general stress-energy tensor components, which we used here.}. The calculated electric field, charge density, and pressure are shown in Fig.~\ref{fig:chargeFieldMetric}(a), and the time and radial component of the resulting metric are shown in Fig.~\ref{fig:chargeFieldMetric}(b). Since $\sqrt{k_J}$ sets the length scale of the problem, we define a unitless measure of distance $\overline{r}=r/\sqrt{k_J}$. 

Due to the electric field, the pressure deviates from Eq.~\ref{tolmanSoln} and passes through zero at $\overline{r}=1.273$. The deviation of $\sqrt{8\pi k_J}r \rho$ and $16\pi r^2 P$ from 1 gives an idea of the length scales at which the electric field becomes important, and can no longer be neglected as $r$ increases. The time component of the metric is arbitrary up to a multiplicative constant, which would be set by asymptotic conditions as $r\rightarrow\infty$, in connecting to an external metric at some finite $r$, or is just an arbitrary scaling factor if our solution describes all space-time. The metric components diverge as $\overline{r}$ approaches about $1.364$. 

The scalar curvature (Ricci scalar) never approaches 0 before that point, so the coordinates are not asymptotically flat. One may attempt to connect the internal metric to an external metric at some $\overline{r}<1.364$\cite{tolman1939}; in this case, matching to the Reissner-Nordstrom metric\cite{reissner1916,nordstrom1918} would be appropriate. However, $\rho$ never approaches zero before the metric diverges, and any connection to a free-space metric (where $\rho$ is set to zero) would necessarily make the pressure discontinuous at the boundary, which violates conservation of the stress-energy tensor. Therefore, no asymptotically flat, spherically symmetric, static solutions appear to exist.

\begin{figure}[h]
{
\includegraphics[scale=0.65]{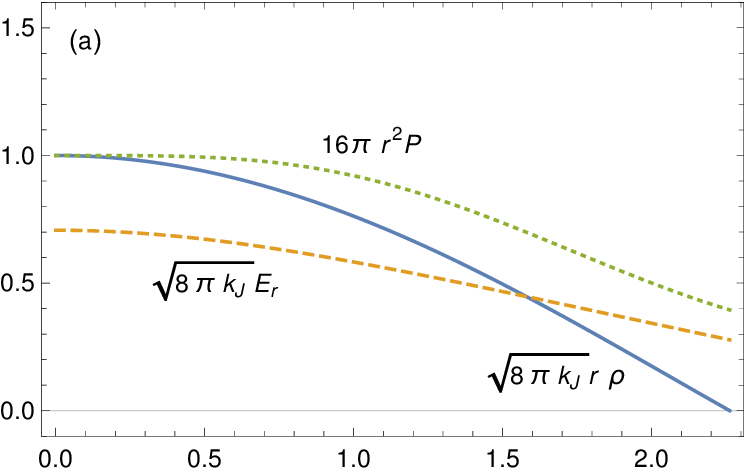}\\
\includegraphics[scale=0.65]{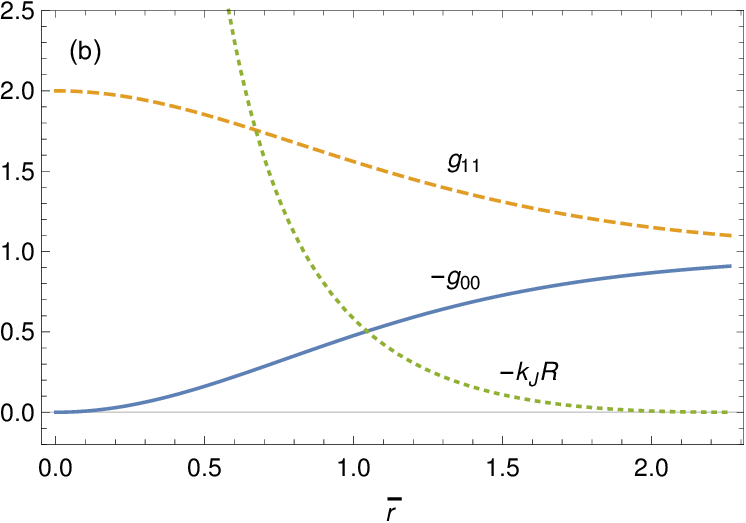}
\caption{\label{fig:chargeFieldMetricNegEM} For $k_{\rm EM}=-1$: (a) the charge density ($\rho=\sqrt{|g_{00}|}J^0$, solid blue curve), radial electric field ($E_r=\sqrt{|g_{00}|g_{11}}F^{01}$, dashed orange curve), and pressure ($P=T_1^1$, dotted green curve) as a function of $\overline{r}$; (b) the time ($-g_{00}$, solid blue curve) and radial ($g_{11}$, dashed orange curve) metric components. The variables have been scaled such that all curves are independent of $k_J$, and $\rho$ and $P$ are shown scaled to the value they would have with no electric field. $g_{00}$ is arbitrary to a multiplicative constant, but all other curves do not depend on this constant. We also show $k_J R$, the scalar curvature to demonstrate that the solutions are singular near the origin.}
}
\end{figure}

With a slight change to the theory, though, we can find an asymptotically flat, stable solution: by changing the sign of $k_{\rm EM}$, which makes the gravitational mass of the electromagnetic field negative. With $k_{\rm EM}=-1$ (again treating $j^\mu$ as independent resulting in $T_J^{\mu\nu}$ as the addition to the stress-energy tensor), using the same method of numerical integration produces solutions for $\rho$, $E_r$ and the metric components shown in Fig.~\ref{fig:chargeFieldMetricNegEM}.  

\begin{figure}[h]
\includegraphics[scale=0.65]{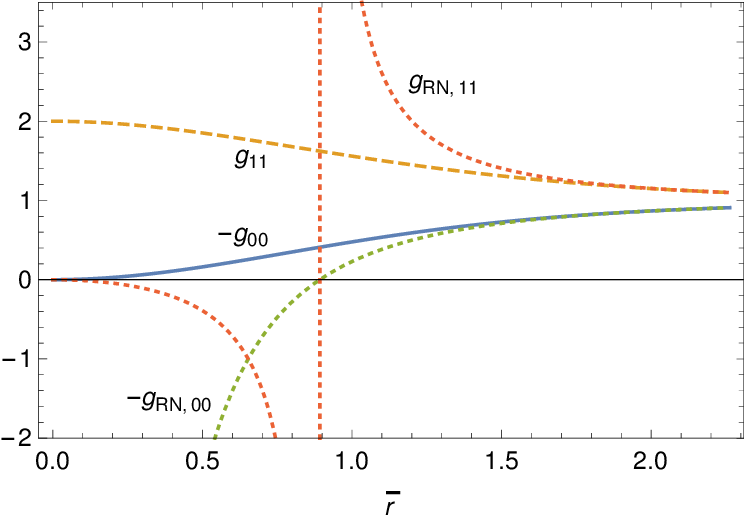}
\caption{\label{fig:rnFit} (a) For $k_{\rm EM}=-1$: the time ($-g_{00}$, solid blue curve) and radial ($g_{11}$, dashed orange curve) metric components; and the Reissner-Nordstrom metric components ($-g_{{\rm RN},00}$ in dotted green and $g_{{\rm RN},11}$ in dotted red with the apparent singularity at $\overline{r}\approx0.9$). $r_1$ has been set appropriately to make the 00 components match at the maximum $\overline{r}$.
  }
\end{figure}

\begin{figure}[h]
\includegraphics[scale=0.65]{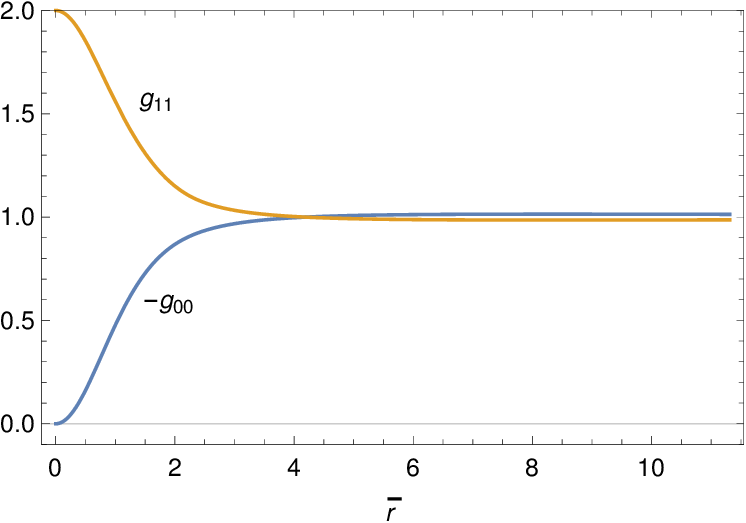}
\caption{\label{fig:rnMetric} For $k_{\rm EM}=-1$: the time ($-g_{00}$, solid blue curve) and radial ($g_{11}$, solid orange curve) metric components; the fact that $g_{11}<1$, $-g_{00}>1$ for large $r$ is a demonstration of the negative mass contained in this distribution.
}
\end{figure}

$\rho$ approaches zero at about $\overline{r}=2.262$, and at that point, the metric is well approximated by the Reissner-Nordstrom metric (but with the sign on the electromagnetic term flipped due to the change of sign of $k_{\rm EM}$),
\begin{equation}
  \begin{array}{rcl}
ds^2 & = & -f_{\rm RN} dt^2 + \frac{1}{f_{\rm RN}} dr^2+r^2d\theta^2+r^2\sin^2\theta d\phi^2 \\
f_{\rm RN} & \equiv & 1-\frac{2m}{r}-\frac{q^2}{8 \pi r^2},
\end{array}
\label{rnNegMetric}
\end{equation}
where $q\equiv 4 \pi r^2 F^{01}$, as measured in the far-field, is the experimentally measured charge. This fit is shown in Fig.~\ref{fig:rnFit}, with $m=-0.1206\sqrt{k_J}$, and $q=3.564\sqrt{k_J}$. 
The conversion of these numbers to SI units is,
\begin{equation}
  \begin{array}{rcl}
    m_{\rm SI} & = & \frac{c^2}{G} m \\
    q_{\rm SI} & = & c^2\sqrt{\frac{\epsilon_0}{G}} q,
  \end{array}
\label{mqToSI}
\end{equation}
yielding $m=-1.62\times 10^{26}{\rm \frac{kg}{m}}\sqrt{k_j}$ and $q_{\rm SI}=1.17\times10^{17}{\rm \frac{kg}{m}}\sqrt{k_J}$. We obtain the fundamental charge if we set $\sqrt{k_J}=1.37\times 10^{-36}{\rm m}$, which is quite a compact charge, but with a mass of about 20 orders of magnitude too high, if we're talking about an electron, not to mention that the gravitational mass is negative. In fact, a negative gravitational mass for a charged particle in this theory with $k_{\rm EM}=-1$ is what one would want. We've made the gravitational electromagnetic energy negative. If that's the case, then for positive charges to accelerate in the direction of the electric field, and negative charges against it, the gravitational mass of the charged particle must have the same sign as the electromagnetic field. If one inspects the long-range behavior of the metric, this negative mass is apparent, as shown in Fig.~\ref{fig:rnMetric}; note how $g_{00}$ and $g_{11}$ cross through 1.

This solution and the theory it comes from are of limited utility, since they don't include some of the more interesting terms (it doesn't violate parity)\footnote{Also, note that the solution here is only a solution of Einstein's equations and the energy-momentum equations, {\em not} the field equations, Eq.~\ref{fieldEqnsFromJ}, since the field equations require $A_\mu=0$ where $J_\mu=0$. This is an example where the solution to the diffeomorphism invariant theory is not a solution to the associated least action theory.} and to find a solution we had to make electromagnetic energy negative. It's just the simplest toy model we can construct that self-consistently deals with extended charged objects. While it predicts a charge to mass ratio that is far from anything resembling any of the fundamental particles, the fact that we get numbers at all is an exciting prospect. This theory predicted a single mass and charge for a stable object; the predicted mass and charge {\em are quantized}. The singularity at the origin prevented us from setting anything about the central density of the charge (which is what is done in non-singular models such as this in general relativity\cite{bonazzola1993}); there is only one solution, with a set mass and charge. To the author's knowledge, this is the only time classically that this has been demonstrated by any theory. 

It's also worth noting that static, spinning perfect fluid solutions (in general relativity) have been found in situations of very high central energy density, where the pressure approaches proportionality to the energy density, and these solutions demonstrate this same behavior, since the coordinates approach spherical symmetry near the object's center\cite{flammer2018}. This singular behavior is quite interesting, because it removes one degree of freedom from the solution space. If another such condition exists on the angular momentum of the solutions, spin would be quantized as well. The author assumes the parity violating term could play a role in that respect.

\subsubsection{\label{sec:solutionAConst}Lagrangian of Eq.~\ref{matterLagrangian} with $A_\mu$ as Independent}

Now let us turn our attention to static, spherical solutions to the next simplest theory, that of the Lagrangian of Eq.~\ref{matterLagrangian}, but varying $A_\mu$ rather than $j^\mu$. For our static, spherical situation, we can simplify the stress-energy tensor, $T_{J,A}^{\mu\nu}$ significantly. By our spherically symmetric, static assumption, the only components of $F^{\mu\nu}$ are $F^{01}=-F^{10}$, which only depend on $r$. Under these conditions, the field equations, Eq.~\ref{fieldEqnsFromA}, can be integrated to obtain the following relationship,
\begin{equation}
\begin{array}{rcl}
  \nabla^{[\mu}J^{\nu]} & = & -\frac{k_{\rm EM}}{2 k_J}F^{\mu\nu} + \frac{1}{k_J}C^{\mu\nu} \\
  C^{01} & = & - C^{10} = \frac{1}{\sqrt{|g_{00}g_{11}|}r^2}C_0, 
\end{array}
\label{fieldEqnIntegralSpherical}
\end{equation}
where $C^{\mu\nu}$ is a container for the integration constant $C_0$ (all components of $C^{\mu\nu}$ are zero outside of $C^{10}$ and $C^{01}$). Using this relationship, we can rewrite $T_{J,A}^{\mu\nu}$ for our situation as
\begin{equation}
\begin{array}{rcl}
  T_{J,A}^{\mu\nu} = T_J^{\mu\nu} + 2 k_{\rm EM} T_{\rm EM }^{\mu\nu} + C^{\alpha\beta}F_{\alpha\beta}g^{\mu\nu} + 4 C^{\alpha(\mu}F^{\nu)\beta}g_{\alpha\beta}.
\end{array}
\label{tMuNuJASpherical}
\end{equation}
The total stress energy tensor for our situation is 
\begin{equation}
\begin{array}{rcl}
  T^{\mu\nu} = T_{J,A}^{\mu\nu} - k_{\rm EM} T_{\rm EM }^{\mu\nu},
\end{array}
\label{tMuNuSpherical}
\end{equation}
where $k_{\rm EM}=-1$ to give the conventional sign for electromagnetic energy-momentum. The extra terms in $T_{J,A}^{\mu\nu}$ in addition to $T_J^{\mu\nu}$ (the perfect fluid stress-energy tensor) is minus twice $T_{\rm EM}$ and, in effect, flip its sign! If we set $C_0=0$, we obtain the same stress-energy tensor as in Sec.~\ref{sec:solutionJConst}, where we used negative electromagnetic energy to find a solution. Here, we can use positive  electromagnetic energy, but the additional terms in $T_{J,A}^{\mu\nu}$ negate it, and will provide the necessary binding to produce static solutions. 

It's worth noting that the solution of the last section from $0<\overline{r}<\infty$, where we connected to the Reissner-Nordstrom metric, is not a solution to the field equations or energy-momentum conservation equations if we're treating $A_\mu$ as independent for all $\overline{r}$. The discontinuity in the derivative of $\rho$ at the edge of the distribution creates an infinite value in the conservation of energy-momentum equation in our current situation, due to the higher order derivative of $J^\mu$ that appears in our current stress-energy tensor.

\begin{figure}[h]
\includegraphics[scale=0.65]{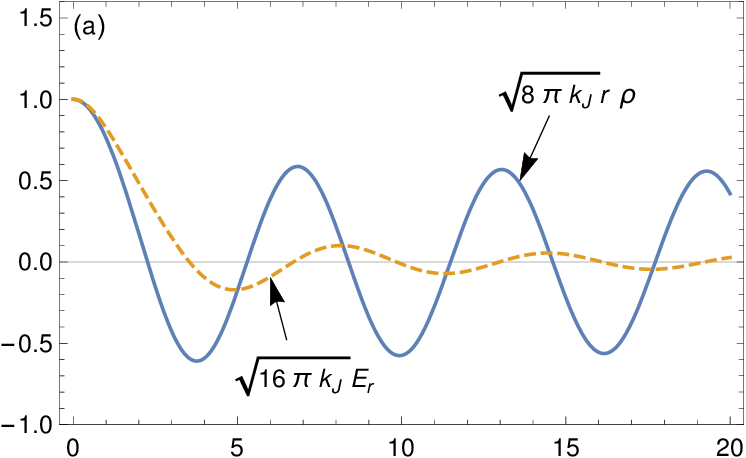} \\
\includegraphics[scale=0.65]{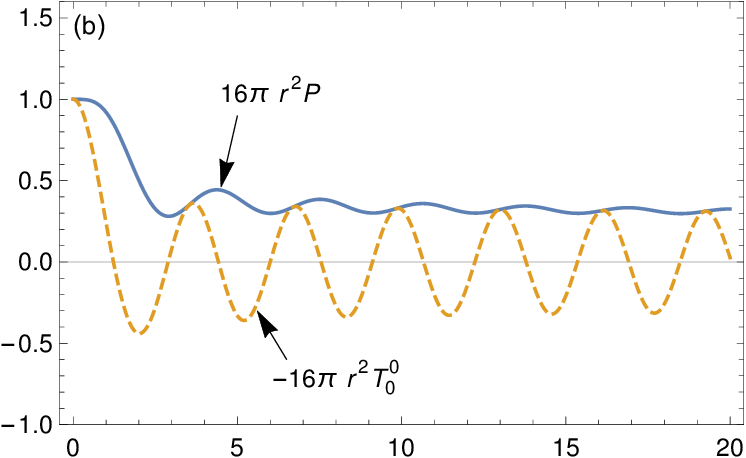} \\
~~\includegraphics[scale=0.63]{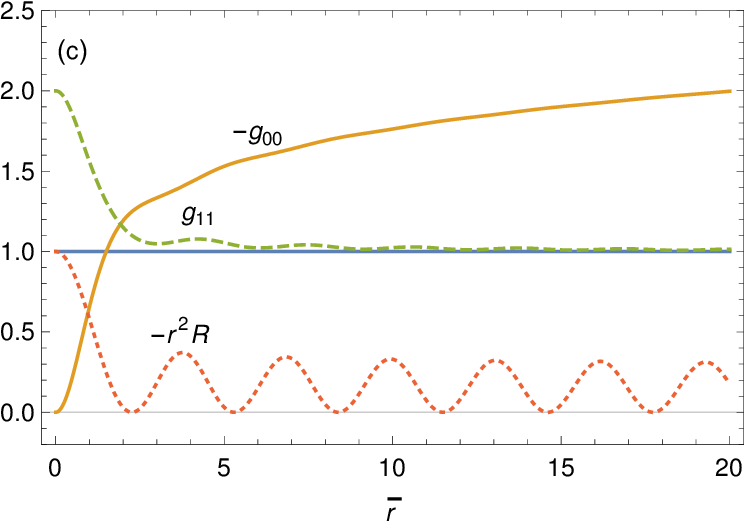}
\caption{\label{fig:AConstSoln} (a) The charge density ($\rho=\sqrt{|g_{00}|}J^0$, solid blue curve), radial electric field ($E_r=\sqrt{|g_{00}|g_{11}}F^{01}$, dashed orange curve); the pressure ($P=T_1^1$, solid blue curve) and $T_0^0$ (dashed orange curve); and (c) the time ($-g_{00}$, solid orange curve) and radial ($g_{11}$, dashed green curve) metric components, and the scalar curvature ($R$, dotted red curve). All are shown as a function of $\overline{r}$ ($r/\sqrt{k_J}$). All curves outside of the metric have added powers of $\overline{r}$ to make them approach 1 at the origin. $g_{00}$ is arbitrary to a multiplicative constant, but all other curves do not depend on this constant.
}
\end{figure}

The solution up to where $\rho$ went to zero in Fig.~\ref{fig:chargeFieldMetricNegEM}, however, is a solution to the energy-momentum conservation equations; we also plugged this solution into the field equations, Eq.~\ref{fieldEqnsFromA}, and within numeric error, these were also satisfied, validating our approach. We can extend this solution beyond where $\rho$ goes to zero. Numerically integrating Einstein's equations out to $\overline{r}=20$, we obtain the fields and metric in Fig.~\ref{fig:AConstSoln} (the solution beyond where $\rho$ first goes to zero still, within numeric error, satisfies the field Equations, Eq.~\ref{fieldEqnsFromA}, for all calculated $\overline{r}$).
Now, around the compact, central sphere of charge are spherical shells of alternating charge. While $\rho$ trends like an oscillating function times $1/r$, the volume in each half-cycle of the oscillation trends as $r^2$. This means the charge enclosed in each half-cycle increases linearly as a function of $r$. $E_r$ still tends to zero, but also trends as an oscillating function times $1/r$. This is interesting considering the dual particle-wave nature of the fundamental particles of nature. It raises the question of whether the dynamics of such a particle-wave as in Fig.~\ref{fig:AConstSoln} can reproduce the dynamics that have been observed in fundamental particles, consistent with quantum mechanics\footnote{Recall that no local equations of motion have existed before now in the literature, so treating such extended charged objects has not been possible.}.

\begin{figure}[h]
~~~~\includegraphics[scale=0.58]{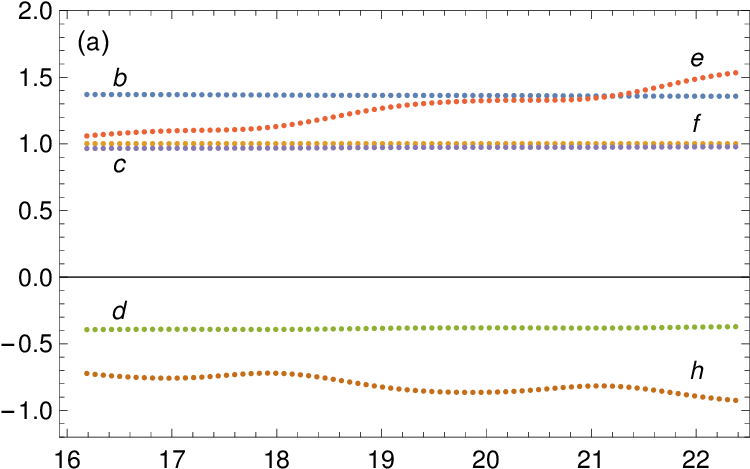}\\
\includegraphics[scale=0.65]{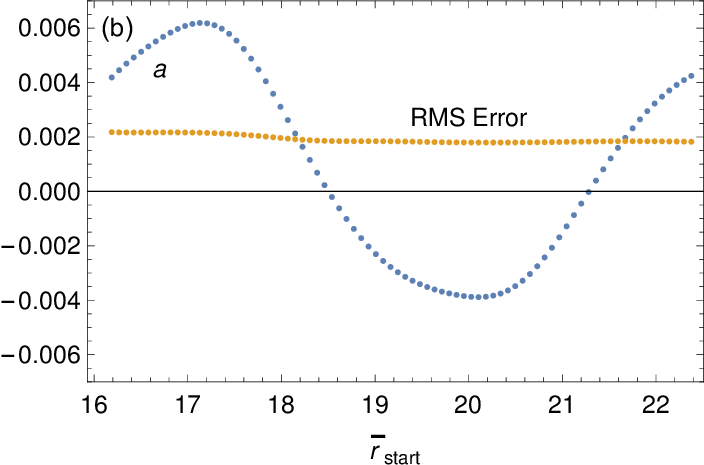}
\caption{\label{fig:AConstFieldFitConsts} (a) $b$, $c$, $d$, $e$, $f$, and $h$ fit constants used in fitting to the charge enclosed, Eq.~\ref{qEncFit}; (b) fit constant $a$ shown with RMS error. Both are fits over 4 periods in $\overline{r}$ (a range of $\Delta \overline{r} = 25.07$), and the horizontal axis is the starting $\overline{r}$ for the range of the fits. 
}
\end{figure}

We've already described the short-range behavior of this solution in the previous section. The long-range behavior for the metric fits well to the following functions\footnote{For $g_{11}$, we fit $10<\overline{r}<50$. For $g_{00}$, one must go out further out to isolate the logarithmic growth, so we fit $50<\overline{r}<100$.},
\begin{equation}
\begin{array}{rcl}
  g_{00} & = & -\frac{1}{r_0^2}(0.817 + 0.375\ln(\overline{r})) \\
  g_{11} & = & \frac{1}{1-2 m_a(\overline{r})/\overline{r}} \\
  m_a(\overline{r}) & = & 0.104 + \frac{0.0350}{\overline{r}} \\
   & + & \left(0.0354 + \frac{0.0591}{\overline{r}}\right)\cos(2.01\overline{r}+3.74).
\end{array}
\label{metricFit}
\end{equation}
The charge enclosed as a function of radius is related to the radial electric field as
\begin{equation}
\begin{array}{rcl}
  \frac{q_{\rm enc}}{\overline{r}} & = & \int\int \sqrt{g_{22}g_{33}} \frac{E_r}{\overline{r}} d\theta d\phi \\
   & = & 4 \pi k_J \overline{r}  E_r,
\end{array}
\label{qEncDef}
\end{equation}
which fits well to
\begin{equation}
  \begin{array}{rcl}
    \frac{q_{\rm enc}}{k_J\overline{r}} & = & \frac{a}{\overline{r}} + b \sin\left(c \overline{r}+d\right) + \frac{e}{\overline{r}}\sin\left(f \overline{r}+h\right) \\
  \end{array}
  \label{qEncFit}
\end{equation}
where $a$ represents what would likely be measured as the net charge of the distribution. Extracting this net charge is somewhat of a challenge. While we can find a fit that is quite good, the electric field at larger $\overline{r}$ is dominated by the oscillations in $\rho$, and the net charge is either small or perhaps zero; $r E_r$ closely oscillates symmetrically around 0. This is demonstrated by trying to fit this curve; some of the constants, including $a$ depend heavily on the range of the fit, as shown in Fig.~\ref{fig:AConstFieldFitConsts}. While the RMS error stays roughly constant, and the parameters that are associated with the larger portions of the fit are quite stable, $a$ in particular varies quite a lot, changing sign depending on the range of the fit chosen. The value of $a$ does tend to be more positive than negative as the range is varied, but its contribution is small compared to the error in the fit.

The mass of this object is also difficult to determine. While the coordinates are asymptotically flat in the sense that the curvature tends to zero as $r\rightarrow\infty$, the metric does not approach a Schwarzschild-like metric as $r\rightarrow\infty$ (due to the logarithmic growth of $g_{00}$), so we cannot extract the mass from the $m$ parameter from such an asymptotic fit. It is interesting to note that if particles are of this nature, their gravitational effect out to $\infty$ is different than a typical Schwarzschild-like mass, and this could have cosmological consequences.

In any case, this solution exhibits the same quantized behavior as the solution of Fig.~\ref{fig:chargeFieldMetricNegEM}. There are no degrees of freedom in the solution that would lead to a continuum of mass/charge. Whatever it is, it is set.

\begin{figure}[h]
\includegraphics[scale=0.65]{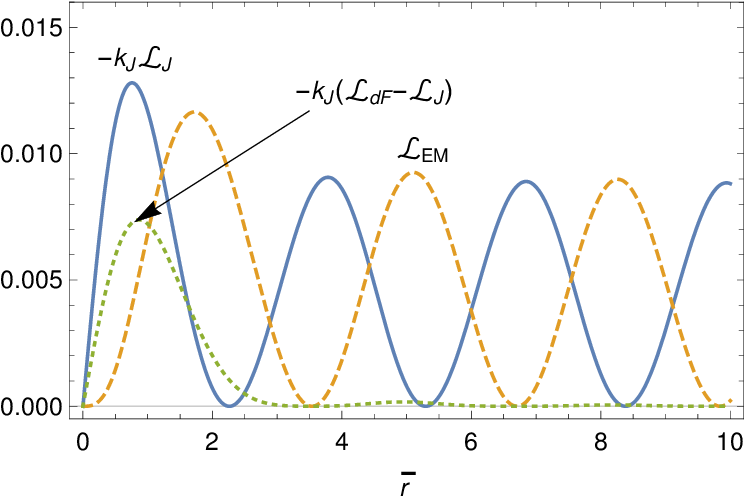}
\caption{\label{fig:lagrangians} For the solution of Fig.~\ref{fig:AConstSoln}, the magnitudes of the different Lagrangian densities to give an idea of their relative magnitudes and ranges. Note for spherical symmetry, $\mathcal{L}_{d*F}=0$.
}
\end{figure}

These simple theories don't include the two curvature-mediated Lagrangians. To demonstrate the relative strengths and ranges of the different Lagrangian components, we plot the values of the different Lagrangian densities for the solution of Fig.~\ref{fig:AConstSoln} as functions of $\overline{r}$ in Fig.~\ref{fig:lagrangians} (evaluated at $\theta=\pi/2$).
For these spherically symmetric coordinates, $\mathcal{L}_{d*F}=0$ identically, which isn't surprising since spherical symmetry doesn't support any kind of parity violation. $\mathcal{L}_{dF}$ has a term that's proportional to $\mathcal{L}_J$, so we subtract that off, $\mathcal{L}_{dF}-\mathcal{L}_{J}=-\frac{1}{4}W_{\alpha\beta\mu\nu}F^{\alpha\beta}F^{\mu\nu}$. This term, due to its dependence on the curvature which approaches singularity at the origin has the same functional dependence near the origin as $\mathcal{L}_J$; it could have an effect on modifying the singular behavior of the solution. The strength of the electromagnetic interaction, represented by $\mathcal{L}_{\rm EM}$, is small compared to the other two terms near the origin, and will have negligible effect, as already demonstrated. $\mathcal{L}_J$ is stronger than $\mathcal{L}_{\rm EM}$ for small $\overline{r}$, but as the gravitational effects die out for larger $\overline{r}$, it approaches the same strength, giving rise to the wave-like structure around the central charge (for a truly compact structure, it would be zero outside the charge distribution due to its direct dependence on $J^\mu$). The curvature dependent $\mathcal{L}_{dF}-\mathcal{L}_{J}$ decays much more rapidly as $\overline{r}$ increases, and is thus quite short in range. $\mathcal{L}_{\rm EM}$ of course has its long range behavior (this would be true even for solutions where $J^\mu$ is zero outside of some boundary), as long as there were some excess charge.

\begin{figure}[h]
\includegraphics[scale=0.65]{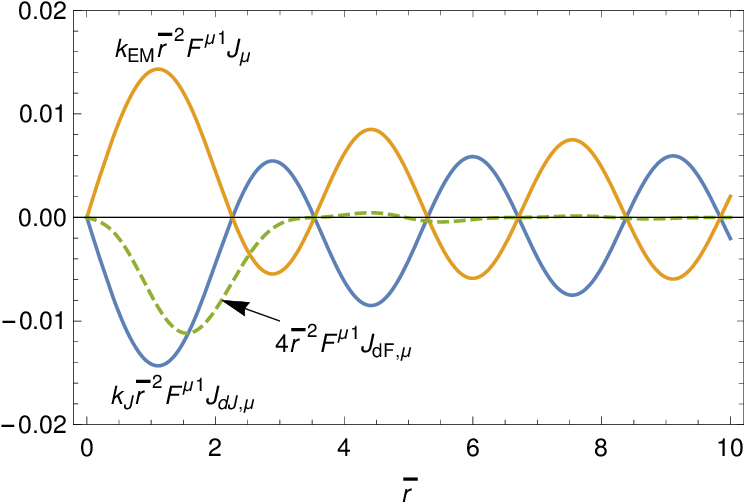}
\caption{\label{fig:eCons} For the solution of Fig.~\ref{fig:AConstSoln}, energy-momentum conservation terms that appear from the more general theory of Eq.~\ref{completeMatterL}, to gauge their relative strength and range.
}
\end{figure}

It could also be interesting to look at how different terms from Eq.~\ref{eConsEquationsOfJs} could play a role in energy-momentum conservation.  This is shown in Fig.~\ref{fig:eCons}.
The two terms that are actually present in the theory that we solved are $F^{\mu 1}J_{dJ,\mu}$ and $F^{\mu 1}J_\mu$, and they are equal and opposite demonstrating energy-momentum conservation. The term involving the curvature-dependent current $J_{dF}^\mu$ has an even weaker functional form near the origin than the other terms, and demonstrates its short-range behavior for large $r$ as well. 

For the first time, we have a theory that can be solved for stable, charged objects. The results are quite promising. Charge and mass quantization have been demonstrated; for a given theory it appears at most one static, spherically symmetric solution will exist. Two new short range forces are present, where one changes sign under a parity transformation. For non-spherical (i.e. spinning) solutions, the parity violating term could require angular momentum quantization in addition to the mass/charge quantization we've already found in spherical symmetry. 

\section{Summary}

We reviewed the history and problems associated with extended-body electromagnetic theory, and the pathologies associated with electromagnetic point charges. These are due the fact that $\nabla_\mu T_{\rm EM}^{\mu\nu}$ is non-zero in the presence of charge, but without point charges, it's been unclear how to complete the stress-energy tensor. Point charges bring their own problems in the form of infinite self-energies, requiring renormalization and perturbative methods to deal with self-interaction; they also are, to-date, incompatible with general relativity.

We then developed self-consistent extended body theories. Initially this was using the fact that in order for a theory to be self-consistent, we need some addition to $T_{\rm EM}$ that maintains only 3 independent energy-momentum conservation equations (to match to the degrees of freedom in the electromagnetic field). Next, we used variational methods. If we remove point charges from the theory, and include in the Lagrangian all quadratic terms involving $F^{\mu\nu}$ and its first derivatives, Eq.~\ref{completeMatterL}, we find terms which, when applying the principle of least action, Eq.~\ref{deltaActionEqualsZero}, result in self-consistent field equations, the solutions of which will satisfy $\nabla_\mu T^{\mu\nu}=0$. One of the terms represents direct current-current interactions, and is stronger than the electromagnetic force at short ranges. For spherical solutions we demonstrated it contributes negative energy and can negate the electromagnetic energy. This provides a mechanism for negative bare mass, which is required for renormalization of very compact charges. The two other additional terms in the Lagrangian of Eq.~\ref{completeMatterL} represent short-range forces, which will be negligible outside of areas of strong curvature (far from the center of a dense, compact object). One of these terms changes sign under a parity transformation, violating parity, as the weak interaction does.

We found static, spherically symmetric solutions using the simplest possible theories, which exclude the short-range, curvature mediated terms. An integrable singularity at the center of the object results in no available degrees of freedom to set the mass or charge: there is exactly one solution with quantized mass and charge. Non-spherical, spinning solutions with quantized mass, charge, and angular momentum may exist. Development of the appropriate computational methods to solve Einstein's equations with the more complicated stress-energy tensor and field equations will be necessary to explore this.

\section{Acknowledgments}
The author gratefully acknowledges fruitful conversations with Dr. Travis Kopp concerning the mathematics contained in this paper.




\end{document}